\documentclass[fleqn,usenatbib]{mnras}

\usepackage{newtxtext,newtxmath}

\usepackage[T1]{fontenc}

\DeclareRobustCommand{\VAN}[3]{#2}
\let\VANthebibliography\thebibliography
\def\thebibliography{\DeclareRobustCommand{\VAN}[3]{##3}\VANthebibliography}

\usepackage{graphicx}	
\usepackage{amsmath}	
\usepackage{wasysym}
\usepackage{xcolor}
\makeatletter 
  \patchcmd{\NAT@citex}
    {\@citea\NAT@hyper@{%
      \NAT@nmfmt{\NAT@nm}%
      \hyper@natlinkbreak{\NAT@aysep\NAT@spacechar}{\@citeb\@extra@b@citeb}%
      \NAT@date}}
    {\@citea\NAT@nmfmt{\NAT@nm}%
    \NAT@aysep\NAT@spacechar\NAT@hyper@{\NAT@date}}{}{}

  \patchcmd{\NAT@citex}
    {\@citea\NAT@hyper@{%
      \NAT@nmfmt{\NAT@nm}%
      \hyper@natlinkbreak{\NAT@spacechar\NAT@@open\if*#1*\else#1\NAT@spacechar\fi}%
        {\@citeb\@extra@b@citeb}%
      \NAT@date}}
    {\@citea\NAT@nmfmt{\NAT@nm}%
    \NAT@spacechar\NAT@@open\if*#1*\else#1\NAT@spacechar\fi\NAT@hyper@{\NAT@date}}
    {}{}
\makeatother

\title[X-ray gas in IllustrisTNG ETGs]{X-ray scaling relations of early-type galaxies in IllustrisTNG and a new way of identifying backsplash objects}

\author[Y. Wang et al.]{Yunchong Wang$^{1,2}$\thanks{E-mail: \url{ycwang19@stanford.edu}},
Mark Vogelsberger$^{3,4}$,
Dong-Woo Kim$^{5}$,
Josh Borrow$^{3}$,
Aaron Smith$^{5}$,
\newauthor
Lars Hernquist$^{5}$
and Wenjie Lin$^{6}$
\\%
\\%
$^{1}$Physics Department, Stanford University, 382 Via Pueblo Mall, Stanford, CA 94305, USA \\%
$^{2}$Kavli Institute for Particle Astrophysics \& Cosmology, P. O. Box 2450, Stanford University, Stanford, CA 94305, USA \\%
$^{3}$Kavli Institute for Astrophysics and Space Research, Department of Physics, MIT, Cambridge, MA 02139, USA \\%
$^{4}$ The NSF AI Institute for Artificial Intelligence and Fundamental Interactions, Massachusetts Institute of Technology, Cambridge, MA 02139, USA \\%
$^{5}$Center for Astrophysics $\vert$ Harvard $\&$ Smithsonian, 60 Garden Street, Cambridge, MA 02138, USA \\%
$^{6}$Department of Mechanical Engineering, Columbia University, NYC, NY 10027, USA
}

\date{Accepted ***. Received ***; in original form ***}

\pubyear{2023}

\begin{document}
\label{firstpage}
\pagerange{\pageref{firstpage}--\pageref{lastpage}}
\maketitle
\begin{abstract}
We investigate how feedback and environment shapes the X-ray scaling relations of early-type galaxies (ETGs), especially at the low-mass end. We select central-ETGs from the IllustrisTNG-100 box that have stellar masses $\log_{10}(M_{\ast}/\mathrm{M_{\astrosun}})\in[10.7, 11.9]$. We derive mock X-ray luminosity ($L_{\mathrm{X, 500}}$) and spectroscopic-like temperature ($T_{\mathrm{sl, 500}}$) of hot gas within $R_{500}$ of the ETG haloes using the MOCK-X pipeline. The scaling between $L_{\mathrm{X, 500}}$ and the total mass within 5 effective radii ($M_{5R_{\rm e}}$) agrees well with observed ETGs from Chandra. IllustrisTNG reproduces the observed increase in scatter of $L_{\mathrm{X, 500}}$ towards lower masses, and we find that ETGs with $\log_{10} (M_{5R_{\rm e}}/\mathrm{M_{\astrosun}}) \leqslant 11.5$ with above-average $L_{\mathrm{X, 500}}$ experienced systematically lower cumulative kinetic AGN feedback energy historically (vice versa for below-average ETGs). This leads to larger gas mass fractions and younger stellar populations with stronger stellar feedback heating, concertedly resulting in the above-average $L_{\mathrm{X, 500}}$. The $L_{\mathrm{X, 500}}$--$T_{\mathrm{sl, 500}}$ relation  shows a similar slope to the observed ETGs but the simulation systematically underestimates the gas temperature. Three outliers that lie far below the $L_{\rm X}$--$T_{\rm sl}$ relation all interacted with larger galaxy clusters recently and demonstrate clear features of environmental heating. We propose that the distinct location of these backsplash ETGs in the $L_{\rm X}$--$T_{\rm sl}$ plane could provide a new way of identifying backsplash galaxies in future X-ray surveys. 
\end{abstract}




\section{Introduction}
\label{sec:1}

Scaling relations are fundamental modalities between different physical properties of a certain class of astrophysical objects. Understanding these scaling relations is crucial to understanding the core physics that govern the formation and evolution of these objects. Over the past few decades, scaling relations for galaxy clusters have been studied extensively with a lush set of multiple frequency band observations. These relations mainly relate cluster mass estimates from weak lensing~\citep{2014MNRAS.439....2V,2015MNRAS.446.2205M,2015MNRAS.449..685H} to their electromagnetic signal in the X-ray~\citep{2009ApJ...692.1033V,2010MNRAS.406.1773M,2016MNRAS.463.3582M}, millimeter~\citep{2015ApJS..216...27B,2016A&A...594A..27P,2018ApJS..235...20H} and optical~\citep{2011AJ....142...72E,2014ApJ...785..104R,2016ApJS..224....1R,2018PhRvD..98d3526A} wavelengths. Studying these scaling relations have significantly improved our understanding of galaxy formation at the high-mass end~\citep{1998ApJ...495...80B, 2012ARA&A..50..353K} as well as enabling more precise cluster cosmology with better halo mass estimation~\citep{2010MNRAS.406.1759M,2011ARA&A..49..409A}.

Extending these scaling relations from galaxy clusters to the massive galaxies is both  challenging and important for understanding the formation of massive galaxies. Thanks to recent dedicated efforts targeting early-type galaxies (ETGs, e.g. \citealt{2009ApJS..182..216K}), observers are now starting to study ETG scaling relations at a similarly detailed level as galaxy clusters. It is found that ETG mass and their relatively hot interstellar medium (ISM) and circumgalactic medium (CGM) also follow scaling relations, although in most cases with different slope and scatter from those of groups and clusters. These include the X-ray luminosity--mass relation~\citep{2001MNRAS.328..461O,2006ApJ...653..207D,2011ApJ...729...12B,2013ApJ...776..116K,2013MNRAS.432.1845S,2015ApJ...812..127K,2016ApJ...826..167G,2017MNRAS.464L..26F,2018ApJ...857...32B,2019MNRAS.488.1072K}, X-ray luminosity--temperature relation~\citep{2006ApJ...653..207D,2011ApJ...729...12B,2015ApJ...812..127K,2016ApJ...826..167G,2018ApJ...857...32B,2019MNRAS.488.1072K}, and temperature--mass relation~\citep{2003MNRAS.340.1375O,2016ApJ...826..167G,2018ApJ...857...32B}. The latest data products from Chandra also provide 2D spectral maps of the X-ray gas in ETGs~\citep{2019ApJS..241...36K}, enabling analyses of the spatial distribution and dynamical properties of hot gas in ETGs. Diverse shapes of gas temperature profiles have been found from these X-ray spectral maps which indicate distinct sources of internal (feedback) and external (environment) heating being present for the X-ray gas in ETGs~\citep{2020MNRAS.492.2095K}.

Interestingly, many of these recent studies~\citep{2013ApJ...776..116K,2015ApJ...812..127K,2016ApJ...826..167G,2019MNRAS.488.1072K} also found that the scatter in these ETG X-ray scaling relations tends to increase towards lower masses ($\log_{10} (M_{5R_{\rm e}}/\mathrm{M_{\astrosun}}) \leqslant 11.5$). This indicates potential secondary processes driving the differences in the luminosity and temperature of the ISM and CGM at a fixed mass scale. While \citet{1991ApJ...367..476W} found that galaxies at a fixed $B$-band luminosity often have lower X-ray luminosity in denser environments,  more recent literature \citep[e.g.][]{2016ApJ...826..167G} finds a negligible correlation between the scatter in gas temperature and environment. Since supernova feedback (along with other types of stellar feedback, e.g., \citealt{2003MNRAS.339..289S,2005MNRAS.361..776S,2009ApJ...695..292C}) and active galactic nuclei feedback (AGN, e.g., \citealt{2005Natur.433..604D,2006MNRAS.365...11C,2012ARA&A..50..455F}) can act as heat sources and drive outflows of the ISM and CGM, their interplay could lead to the observed scatter in X-ray luminosity or temperature~\citep{2018ApJ...857...32B}. 

However, the debate is still open in both observational and theoretical fronts on whether stellar or AGN feedback is the dominant source of the low-mass-end scatter for ETG X-ray scaling relations. \citet{2006ApJ...653..207D}, using a heating rate argument, suggested that Supernova Ia (SNIa for short) feedback dominates over AGN feedback in gas heating and driving outflows in low mass ETGs from Chandra. Similarly, \citet{2011ApJ...738...57P} argued for SNIa as the dominant source of internal gas heating using semi-analytic modeling in massive ETGs. Nevertheless, \citet{2015MNRAS.449.4105C} found in cosmological hydrodynamic simulations that a combination of kinetic and thermal AGN feedback can significantly reduce ETG X-ray luminosity at fixed mass. \citet{2017MNRAS.464L..26F} further compared ETGs from the SLUGGS survey to \citet{2015MNRAS.449.4105C} and concluded that AGN is the main secondary factor that affects $L_{\rm X}$ in low-mass ETGs. \citet{2017ApJ...835...15C}, using 2D magneto-hydrodynamic zoom-in simulations of individual ETGs, suggested that SNIa and AGN affect distinct aspects of the ETG gas: while gas heating is dominated by SNIa, AGN feedback significantly reduces $L_{\rm X}$. \citet{2019MNRAS.488.1072K} further supports the stellar feedback-driven heating scenario where Chandra-observed ETGs with higher stellar masses tend to have higher $L_{\rm X}$ at fixed dark matter halo masses. Moreover, \citet{2020MNRAS.492.2095K} found that hot gas cores in Chandra ETGs are mostly related to recent star formation and hence stellar feedback is dominant instead of AGN feedback or gravitational heating. In addition to stellar and AGN feedback, other physical differences such as the inner density profile or the dynamical state~\citep{1999A&A...351..487P,2013MNRAS.432.1845S,2015ApJ...812..127K,2019MNRAS.486.4686K} of the ETG can also lead to systematic differences in X-ray luminosities at the low-mass end, pointing to secondary heating effects from galactic rotation and the shape of the gravitational potential which could also add to the low-mass-end scatter.

Advancements in cosmological simulations over the past two decades have revolutionized our understanding of galaxy evolution (see e.g. \citealt{2020NatRP...2...42V} for a review). To further elucidate the origin of these ETG X-ray scaling relations, we use a legacy ETGs sample~\citep{2020MNRAS.491.5188W} from the cosmological hydrodynamic simulation IllustrisTNG, which is an updated version of the Illustris Project~\citep{2014MNRAS.444.1518V, 2014Natur.509..177V, 2014MNRAS.445..175G, 2015MNRAS.452..575S,2015A&C....13...12N}, and that has well-studied density profile, stellar properties, and dark matter fractions~\citep{2018MNRAS.481.1950L,2019MNRAS.490.5722W,2020MNRAS.491.5188W,2022MNRAS.513.6134W}. Our work further extends the previous X-ray scaling relation studies using IllustrisTNG (e.g., X-ray scaling relations for star-forming and quenched galaxies: \citealt{2020MNRAS.494..549T}; the relation of black hole growth to CGM properties:~\citealt{2020ApJ...893L..24O,2021MNRAS.501.2210T}; X-ray scaling relations in galaxy groups and clusters:~\citealt{2022arXiv220511528P}) by especially focusing on the formation mechanisms leading to the scatter at the low-mass-end. Our work is also complementary to these previous X-ray scaling relation studies as we examine outliers showing clear signs of environmental (shock) heating which has important implications for future X-ray surveys.

This paper is organized as follows. In Section~\ref{sec:2} we describe the general information of the simulation, how the mock ETG sample is selected, and how the mock X-ray properties of the hot gas in these ETG parent haloes are extracted. In Section~\ref{sec:3} we present the key results of our analysis including the X-ray scaling relations as well as the physical processes that contribute to shaping these relations. In Section~\ref{sec:4} we present the formation history for outliers of the X-ray luminosity-temperature relation and how they can be used to identify backsplash objects. Finally in Section~\ref{sec:5}, we provide a brief summary of this work.

\section{Methodology}
\label{sec:2}

\subsection{Simulation overview}
\label{sec:2.1}

{\it The Next Generation} Illustris simulations~\citep{2018MNRAS.480.5113M,2018MNRAS.477.1206N,2018MNRAS.475..624N,2018MNRAS.475..648P,2018MNRAS.475..676S,2019MNRAS.490.3234N,2019MNRAS.490.3196P}, a.k.a. IllustrisTNG, is a suite of magneto-hydrodynamic simulations run with the publicly-available moving-mesh code \textsc{Arepo}~\citep{2010MNRAS.401..791S,2020ApJS..248...32W}. It steps up from the original Illustris Simulations~\citep{2013MNRAS.436.3031V,2014MNRAS.438.1985T,2014MNRAS.444.1518V, 2014Natur.509..177V, 2014MNRAS.445..175G, 2015MNRAS.452..575S,2015A&C....13...12N} and improves upon the sub-grid physics of the stellar and AGN feedback models~\citep{2017MNRAS.465.3291W,2018MNRAS.473.4077P}. These improvements lead to more realistic predictions in terms of observed galaxy properties and demonstrates the capacity of IllustrisTNG to shed light on the underlying physical processes shaping these properties. Some of the comparison works with observations include the galaxy mass--metallicity relation~\citep{2018MNRAS.477L..16T,2019MNRAS.484.5587T}, the galaxy-color bimodality in the Sloan Digital Sky Survey~\citep{2018MNRAS.475..624N}, the intra-cluster metal distribution~\citep{2018MNRAS.474.2073V}, early-type galaxy total density profiles~\citep{2020MNRAS.491.5188W}, gas-phase metallicity gradients in star-forming galaxies~\citep{2021MNRAS.506.3024H}, stellar orbital fraction and outer kinematic structure~\citep{2019MNRAS.489..842X,2022MNRAS.513.6134W}, optical morphologies of galaxies~\citep{2019MNRAS.483.4140R}, the size evolution of galaxies~\citep{2018MNRAS.474.3976G}, star formation activities and quenched fractions~\citep{2019MNRAS.485.4817D}, spatially-resolved star formation in galaxies~\citep{2021MNRAS.508..219N}, the fraction of cool-core clusters~\citep{2018MNRAS.481.1809B}, as well as AGN galaxy occupation and X-ray luminosities~\citep{2018MNRAS.479.4056W,2019MNRAS.484.4413H,2020MNRAS.493.1888T}, and predictions of high redshift galaxy luminosity functions for JWST~\citep{2020MNRAS.492.5167V}. Although certain aspects of these predictions are still discrepant with observations, the broad agreement in many properties related to gas cycle, star formation, and feedback lends us generous predicative power to gain insights on the key factors that shape the X-ray scaling relations in ETGs. The simulation adopts Planck-2016 flat-$\Lambda$CDM cosmology~\citep{2016A&A...594A..13P} with parameters of $h = 0.6774$,  $\sigma_{\mathrm{8}} = 0.8159$, $\Omega_{\mathrm{m}} = 0.3089$, $\Omega_{\mathrm{b}} = 0.0486$, and $\Omega_{\mathrm{\Lambda}} = 0.6911$.

\subsection{Sample selection}
\label{sec:2.2}

We select massive (stellar mass $\log_{10} (M_{\ast}/\mathrm{M_{\astrosun}}) \in [10.7, 11.9]$) early-type central galaxies from the TNG100-1 box. This is the highest resolution simulation box that has a side length of $75\,\mathrm{Mpc}/h$, with $2\times1820^{3}$ resolution elements for baryons and dark matter particles. The mass resolution of baryons and dark matter are  $m_{\mathrm{baryon}} = 1.4\times10^{6}\ \mathrm{M}_{\mathrm{\astrosun}}$ and  $m_{\mathrm{DM}} = 8.9\times 10^{6}\ \mathrm{M}_{\mathrm{\astrosun}}$, respectively. This places the TNG100-1 box at a sweet spot for our purpose among the $35\,\mathrm{Mpc}/h$ and $205\,\mathrm{Mpc}/h$ side length boxes, which provides an abundant sample of well-resolved massive galaxies. The softening scale of gas cells are fully adaptive (minimum $0.19\,\mathrm{kpc}$) while a fixed softening length of $\epsilon = 0.74\,\mathrm{kpc}$ is applied to dark matter and stellar particles. All simulation data used for this analysis is publicly available~\citep{2019ComAC...6....2N}.~\footnote{\url{https://www.tng-project.org/}}

We adopt the early-type galaxy classification method documented in \citet{2017MNRAS.469.1824X} to select our simulated ETGs. The final ETG galaxy sample with 559 galaxies is identical to the one used in \citet{2020MNRAS.491.5188W} and we briefly outline the selection procedure. To begin with, the stellar component of the largest gravitationally bound object (found by \textsc{Subfind}~\citealt{2001MNRAS.328..726S,2009MNRAS.399..497D}) in a Friends-of-Friends (FoF) group is defined as the `central' galaxy. An  age and metallicity-dependent magnitude is assigned to stellar particles based on their intrinsic luminosity using the  stellar population synthesis (SPS) model \textsc{galaxev}~\citep{2003MNRAS.344.1000B}. We also apply a semi-analytic dust attenuation model as in \citet{2017MNRAS.469.1824X} to account for dust absorption, emission, and scattering.  

The main classification criterion for a central galaxy to be early-type is based on their SDSS $r$-band rest-frame luminosity profiles. We perform both single and double-component luminosity profile template fitting to enable more robust classification. The single-component fit consists of fitting either exponential (S$\mathrm{\acute{e}}$rsic $n=1$) or de Vaucouleurs (S$\mathrm{\acute{e}}$rsic $n=4$) profile templates. The double-component fit combines a de Vaucouleurs and an exponential profile that has their relative ratio as a free parameter, which are better for galaxies that demonstrate prominent bulge-disk combinations. In the final sample, we only define galaxies as ETGs when they are not only better fitted by a single de Vaucouleurs profile but also have $>50\%$ bulge ratio from the two-component fit in all three (box $x,y,z$ axes) projections. This leads to a sample of 559 central ETGs in the mass range of $\log_{10}(M_{\ast}/\mathrm{M_{\astrosun}})\in[10.7, 11.9]$ (corresponding to [$3.6\times10^4$, $5.8\times10^5$] stellar particles) at $z=0$. The $M_{500}$ mass range for the host haloes of our selected ETGs is in the range of $\log_{10} (M_{500}/\mathrm{M_{\astrosun}}) \in [11.79, 13.67]$.

\subsection{The mock X-ray luminosity and gas temperature}
\label{sec:2.3}

\begin{figure}
	\includegraphics[width=\columnwidth]{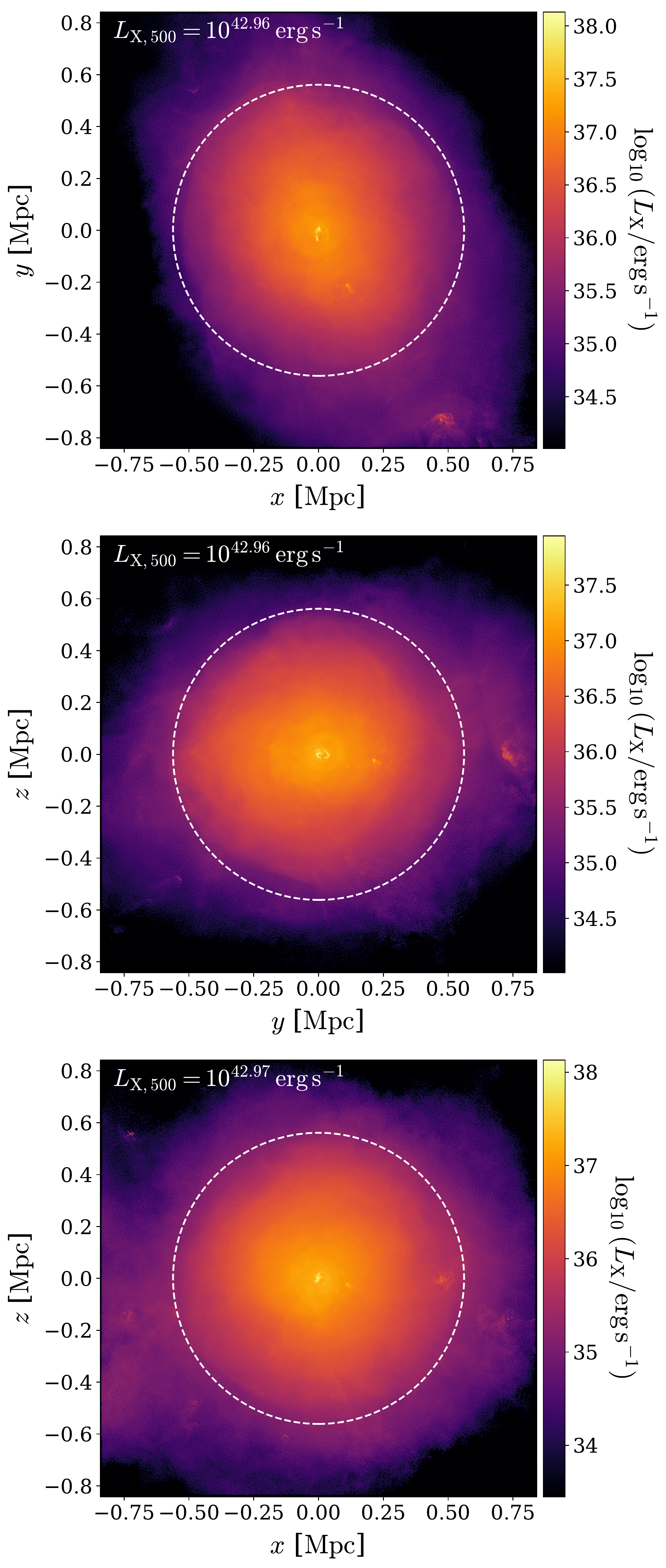}
    \caption{The projected $L_{\mathrm{X}}$ distribution of the hot circumgalactic gas in the host halo of our most massive ETG as a demonstration of the \textsc{MOCK-X} pipeline output. The projections in the $z, x, y$ directions along the simulation box axes are shown from top to bottom. In each panel, the color map indicates the total X-ray luminosity projected in each pixel, while the dashed circle marks the size of the halo $R_{500}$. We also label the total X-ray luminosity projected within a 2D aperture with the size of $R_{500}$ at the top left corner of each subplot. }
    \label{fig:1}
\end{figure}

We create mock X-ray luminosity ($L_{\mathrm{X}}$) maps of the hot circumgalactic medium (CGM) in the parent halo of our selected central ETGs using \textsc{MOCK-X}\footnote{\url{https://bitbucket.org/djbarnes_88/mock-x_public/src/master/}} \citep{2021MNRAS.tmp.1271B}. In each mock pixel, the code starts by generating an X-ray spectral template look-up table for the 11 elements tracked by the simulation in the temperature range $10^{6} < (T/\mathrm{K})< 10^{9}$  (temperature resolution $\delta \log_{10} (T/\mathrm{K}) = 0.02$) using \textsc{apec}~(Astrophysical Plasma Emission Code, \citealt{2001ApJ...556L..91S}) and \textsc{pyatomdb} using atomic data values from \textsc{atomdb} v3.0.9 \citep{2012ApJ...756..128F}. A synthetic X-ray spectrum for each gas cell is then sampled from this numerical table assuming a {\it Chandra} ACIS-I configuration with an energy range of $0.5 < (E/\mathrm{keV}) < 10$ and energy resolution of 150 eV. Cold ($T<10^6$ K), star-forming, or actively cooling gas cells are removed in the calculation. We also apply a correction for the galactic absorption assuming a constant neutral hydrogen column density of $n_{\mathrm{H}} = 2\times10^{20}\ \mathrm{cm}^{-2}$. Finally, the projected X-ray luminosity maps are created along the $x,y,z$ axes of the simulation box within a circular aperture of $3R_{500}$ (3 times the radius within which the mean matter density is 500 times the cosmic critical density).

In Fig.~\ref{fig:1}, we show the projected X-ray luminosity map for the hot gas in the host halo of the most massive galaxy in our ETG sample. From top to bottom, the $x,y,z$ projections are shown respectively. The dashed circle indicates the size of the halo $R_{500}$ (radius within which the average density of the halo is 500 times that of the critical density of the universe). The resolution (0.492 arcsec) and observational band (soft and medium X-ray, 0.5-2.0 keV) of the maps have been set identical to that of Chandra, which we apply to the whole sample of our selected ETGs. In our following analysis, we sum the $L_{\mathrm{X}}$ values for all pixels that fall within $R_{500}$ in the $x$ projection as our $L_{\mathrm{X, 500}}$ measurement. 

As for the X-ray temperature of our massive ETG sample, the minimum temperature of gas cells can fall below $10^{6}$ K ($\sim0.1$ keV), which makes the \textsc{MOCK-X} spectroscopic fitting via APEC table interpolation unreliable (since it was originally designed for cluster scale temperatures). Instead, we implement an approximate X-ray temperature reconstruction for the hot gas in our massive ETGs following the spectroscopic-like temperature definition in \citet{2004MNRAS.354...10M}. Specifically, we construct spectroscopic-like temperature maps with the same projected spatial grid as the $L_{\mathrm{X}}$ maps mentioned above, but with each pixel weighted by the combination of gas density and temperature:
\begin{equation}
    \label{eq:1}
    T_{\mathrm{sl}} = \frac{\int (n_{\rm H}^2 T^{-3/4}) T dV}{\int (n_{\rm H}^2 T^{-3/4}) dV} = \frac{\sum_{i}n_{\mathrm{H}, i}^{2} T_i^{\frac{1}{4}} V_i}{\sum_{i}n_{\mathrm{H}, i}^{2} T_i^{-\frac{3}{4}} V_i}\,,
\end{equation}
where $n_{\mathrm{H}, i}$, $T_{i}$ and $V_{i}$ are the hydrogen number density, temperature, and Voronoi cell volume of the $i$-th gas cell. The summation is conducted over all gas cells that have an overlap (after accounting for their smoothing length) with the pixel in question.To avoid significant bias introduced by cold-dense gas cells to the spectroscopic-like temperature, we remove all gas cells with temperature $T_{i}<0.05\,\mathrm{keV}$ ($\sim 5.8\times 10^{5}\,\mathrm{K}$), number density $n_{\mathrm{H}, i}>0.1$, or star formation rate SFR$>0$ in the calculation of all the spectroscopic-like temperature map calculations. This cut removes gas cells that are interacting with the Equation of State and avoids the $T<0.05\,\mathrm{keV}$ temperature range where spectroscopic-like temperature becomes more inaccurate. We then average over all gas cells with 3D distances $<R_{500}$ to the galaxy center and derive the average halo $T_{\mathrm{sl, 500}}$ for each ETG.

\begin{figure*}
	\includegraphics[width=2\columnwidth]{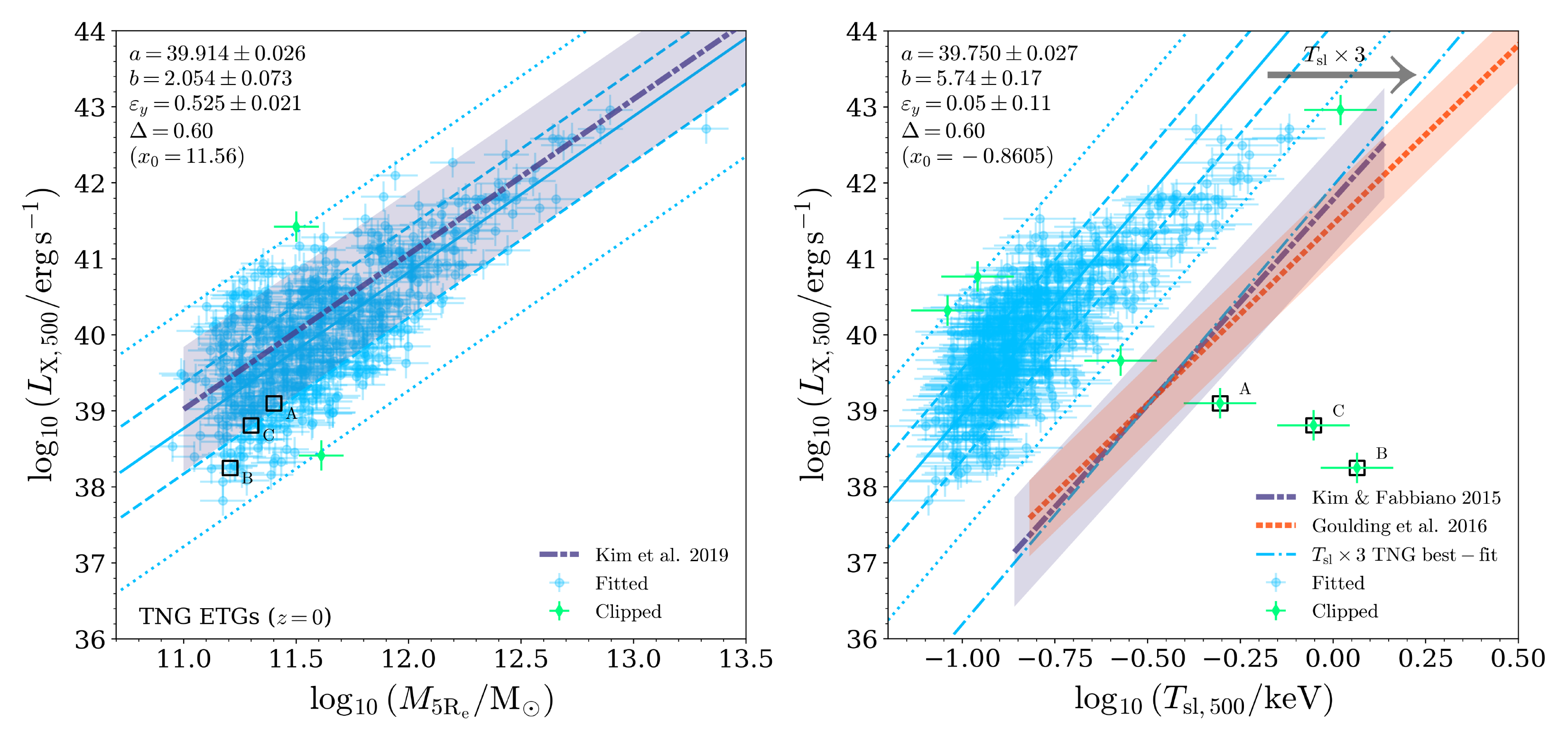}
    \caption{X-ray scaling relations. {\it Left panel:} The total X-ray luminosity projected within $R_{500}$ ($L_{\mathrm{X, 500}}$) versus the total mass within $5R_{\mathrm{e}}$ ($M_{5R_{\rm e}}$) of the central ETG. Colored dots with error bars are IllustrisTNG ETGs with fiducial 0.1 and 0.2 dex uncertainties applied to mass and luminosity mimicking observational uncertainties. The solid, dashed, and dotted blue lines are the best fit scaling relation, $1\sigma$, and $2.6\sigma$ confidence intervals for the simulated ETGs. The blue dots are ETGs within the $2.6\sigma$ ($99\%$) confidence interval, while green dots are outliers. The purple dotted-dashed line and band shows the best-fit scaling relation and the $1\sigma$ confidence interval from Chandra~\citep{2019MNRAS.488.1072K}. The best fit slope ($b$), intercept ($a$), and scatter ($\Delta$) for the IllustrisTNG scaling relation are also labelled in the plot. The three black squares correspond to the three extreme outliers in the right panel. {\it Right panel:} Similar to the left panel, but for the `spectroscopic-like' temperature $T_{\mathrm{sl, 500}}$ versus $L_{\mathrm{X, 500}}$ scaling relation. Fiducial uncertainties of 0.1 dex are applied to the $T_{\mathrm{sl, 500}}$ values. The latest scaling relations for observed ETGs that we compare to are from~\citet{2015ApJ...812..127K,2016ApJ...826..167G}. The three outliers we study in Section~\ref{sec:4} that have significantly higher $T_{\mathrm{sl, 500}}$ than their siblings with similar $L_{\mathrm{X, 500}}$ are marked by the black squares. We also plot dotted-dashed blue the best-fit $L_{\rm X}$--$T_{\rm sl}$ relation for IllustrisTNG ETGs with the temperature multiplied by 3 for a clearer comparison of the scaling relation slope to observations.}
    \label{fig:2}
\end{figure*}

\section{Results}
\label{sec:3}

\subsection{X-ray scaling relations of the simulated ETGs}
\label{sec:3.1}

We present the scaling relation between the X-ray luminosity ($L_{\mathrm{X, 500}}$) and the mass or gas temperature ($T_{\mathrm{sl, 500}}$) of our mock ETGs in Fig.~\ref{fig:2}. We perform linear fit with outlier clipping using the \textsc{lts\_linefit} program\footnote{\url{https://www-astro.physics.ox.ac.uk/~mxc/software/\#lts}} for both scaling relations. The blue dots in the figure mark the mock ETGs that are kept after the clipping, while the outliers (outside of the $2.6\sigma$ confidence interval) are denoted with green dots. We assume a fiducial 0.1 dex uncertainty in the mass and spectroscopic-like temperature measurements, and 0.2 dex uncertainty in the X-ray luminosity.

The left panel shows the scaling relation of $L_{\mathrm{X, 500}}$ with the total mass (dark matter, gas, and stars combined) within five times the 2D effective radii ($R_{\rm e}$) of the central ETGs. There are no significant outliers for our sample and indicates a rather tight linear relation. The dotted-dashed grey line along with the grey band showcases the mean and $1\sigma$ scaling relation fit to Chandra-observed ETGs~\citep{2019MNRAS.488.1072K}. The observers derived ETG dynamical masses using globular cluster (GC) kinematics~\citep{2017MNRAS.468.3949A} or GC photometric scaling relations~\citep{2013ApJ...772...82H,2017ApJ...836...67H} out to $5R_{\rm e}$ and we adopt the same radii for the mass definition of the simulated ETGs. The IllustrisTNG ETGs show great agreement with the observations in terms of both the slope and the scatter of the scaling relation, which is also a significant improvement over past hydrodynamic simulations \citep[e.g.][]{2015MNRAS.449.4105C} thanks to the updated IllustrisTNG AGN and stellar feedback models.  An interesting feature is that the scatter for the simulated ETGs seems to increase towards lower masses, which we investigate further in the following sections (\ref{sec:3.2}, \ref{sec:3.3}).

The right panel of Fig.~\ref{fig:2} shows the scaling relation between $T_{\mathrm{sl}, 500}$ and $L_{\mathrm{X}, 500}$ of the simulated ETGs, which also demonstrates a tight linear relation for most of the points. The two observed ETG samples we compare to are both from integral-field-unit surveys with spatially-resolved stellar kinematics while also being in the Chandra archival footprint. Our best-fit log-linear slope of 5.74 is closer to the 61 $\mathrm{ATLAS^{3D}}$~\citep{2011MNRAS.416.1680C} E and S0 galaxy sample (slope of 5.4, \citealt{2015ApJ...812..127K}) as compared to that of 33 early-type galaxies (slope of 4.7, \citealt{2016ApJ...826..167G}) from the MASSIVE Survey~\citep{2014ApJ...795..158M}. The agreement in the $L_{\rm X}$--$T_{\rm sl}$ slope with observations for a statistical sample of simulated ETGs from a 3D cosmological simulation is also a big step up from previous hydro-simulation results such as  \citet{2017ApJ...835...15C} who used 2D zoom-in approaches. 

Although the slope of the simulated ETGs is close to the observationally-derived values, the mean temperature at fixed $L_{\mathrm{X, 500}}$ for IllustrisTNG is $\sim 0.5$ dex lower than the observed ETGs. Since the $L_{\mathrm{X, 500}}$--$M_{\mathrm{tot}(<5R_{\rm e})}$ relation closely resembles the observed scaling relations, this systematic offset is mainly due to a factor $\sim 3$ colder CGM in IllustrisTNG ETGs (the dotted-dashed blue line in the right panel of Fig.~\ref{fig:2} matches well with observations), that are most likely due to limitations in the current AGN feedback model and will require future improvements to alleviate the discrepancy. \citet{2022arXiv220511528P} found a similar issue with IllustrisTNG (in TNG300) predicting cooler X-ray temperatures of gas than in observations (see their section 4.3 for a discussion on potential model limitations). They found that the spectroscopic temperature derived from MOCK-X in a more massive sample ($M_{500} > 10^{13}\,\mathrm{M_{\astrosun}}$, $T_{\mathrm{X, spec}}>0.5$ keV where spectroscopic template fitting is reliable) is systematically underestimated by a factor $\sim 2$, which is consistent with the $\sim 0.3$ dex underestimation seen in our sample at the high-$T_{\mathrm{sl}}$ end. The spectroscopic-like temperature definition further enhances the underestimation due to the $T^{-3/4}$ weighting, and that underestimation is actually a bit stronger in IllustrisTNG compared to other cosmological hydrodynamic simulations (see Figure A5 in \citealt{2022MNRAS.517.5303L}). 

Despite the cooler-than-observed temperatures for most of our simulated ETGs, there are three significant outliers in $T_{\mathrm{sl, 500}}$ at the low $L_{\mathrm{X, 500}}$ end (Fig.~\ref{fig:2} right panel). Similar high temperature outliers were also seen in observed scaling relations (see e.g. Figure 2 right panel in \citealt{2015ApJ...812..127K}). Although observational uncertainties are large at the faint end, observed ETGs tend to up-scatter more in temperature at fixed low $L_{\rm X}$ ($\sim 10^{38}-10^{39}\ \mathrm{ergs\,s^{-1}}$), which is in the same $L_{\mathrm{X, 500}}$ range as our outliers. Since they are all below the best-fit $L_{\mathrm{X, 500}}$--$M_{5R_{\rm e}}$ relation, we conjecture they are all influenced by galaxy interactions that caused gas stripping. We thoroughly investigate the merger histories of these three outliers in Section~\ref{sec:4} and we identify these outliers as backsplash galaxies that have recently been through significant tidal interactions and environmental heating.

\subsection{The scatter in $L_{\mathrm{X, 500}}$ and the dark matter fraction}
\label{sec:3.2}

\begin{figure*}
	\includegraphics[width=2\columnwidth]{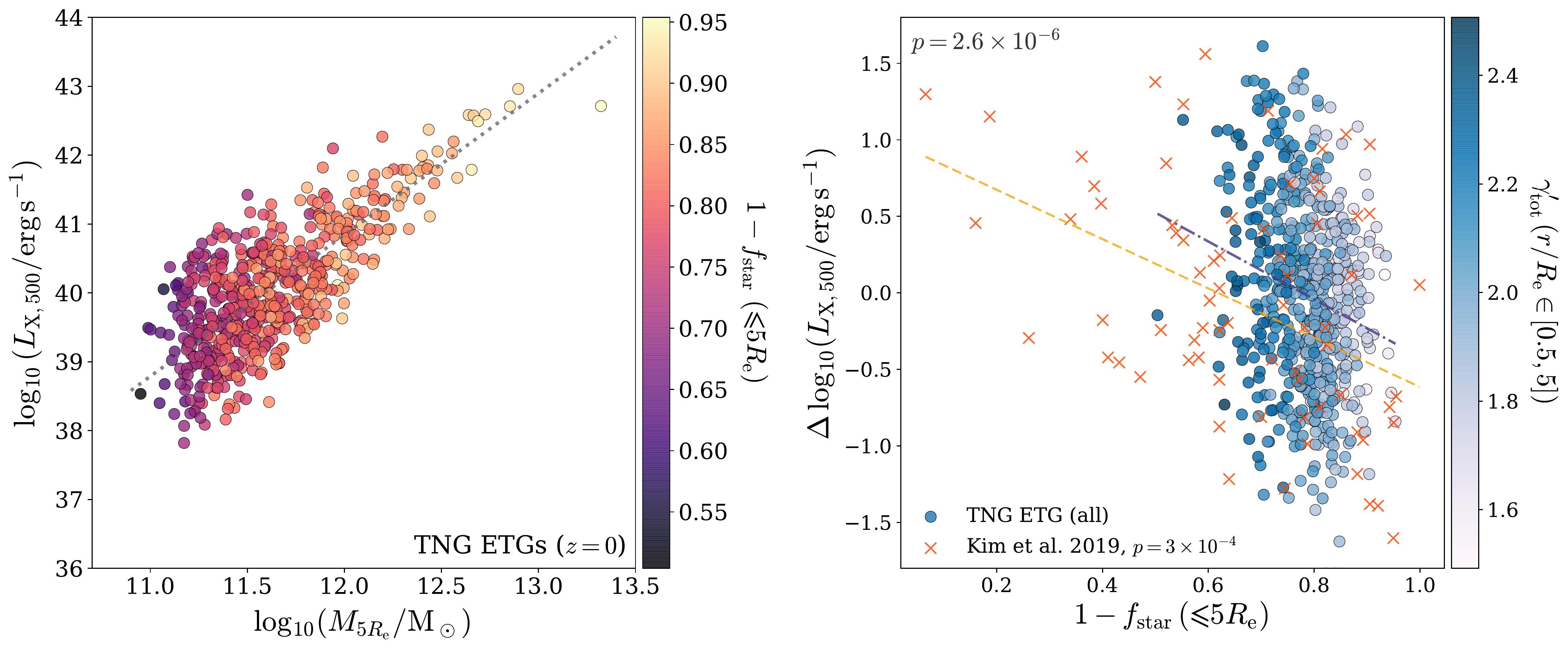}
    \caption{{\it Left panel}: The $L_{\mathrm{X, 500}}$--$M_{5R_{\rm e}}$ scaling relation colored by the combined dark matter and gas mass fraction within $5R_{\rm e}$, $1-f_{\rm star}(\leqslant 5R_{\rm e})$. The grey dashed line denotes the linear fit to the $L_{\mathrm{X, 500}}$--$M_{5R_{\rm e}}$ scaling relation (same as the solid blue line in the left panel of Fig.~\ref{fig:2}) {\it Right panel}: The offset in the $L_{\mathrm{X, 500}}$ from the best-fit $L_{\mathrm{X, 500}}$--$M_{5R_{\rm e}}$ scaling relation, $\Delta L_{\mathrm{X, 500}}$, versus $1-f_{\rm star}(\leqslant 5R_{\rm e})$. The orange markers are observed ETGs from \citet{2019MNRAS.488.1072K}, while the orange dashed line is the linear fit to them. The color map for the scattered points indicates the power-law slope of the simulated ETG total density profiles measured in 3D spherical shells from $0.5R_{\mathrm{e}}$ to $5R_{\mathrm{e}}$. The blue dotted-dashed line is the linear fit to the IllustrisTNG ETGs with a Pearson $p=2.6\times 10^{-6}$, indicating a clear correlation between  $L_{\rm X, 500}$ and $1-f_{\rm star}$ as in observations. However, the simulated ETGs have systematically large dark matter fractions than Chandra ETGs and their $\Delta L_{\rm X}$ show no clear correlation with their density profile slope $\gamma^{\prime}_{\rm tot}$.}
    \label{fig:3}
\end{figure*}

\begin{figure*}
	\includegraphics[width=2\columnwidth]{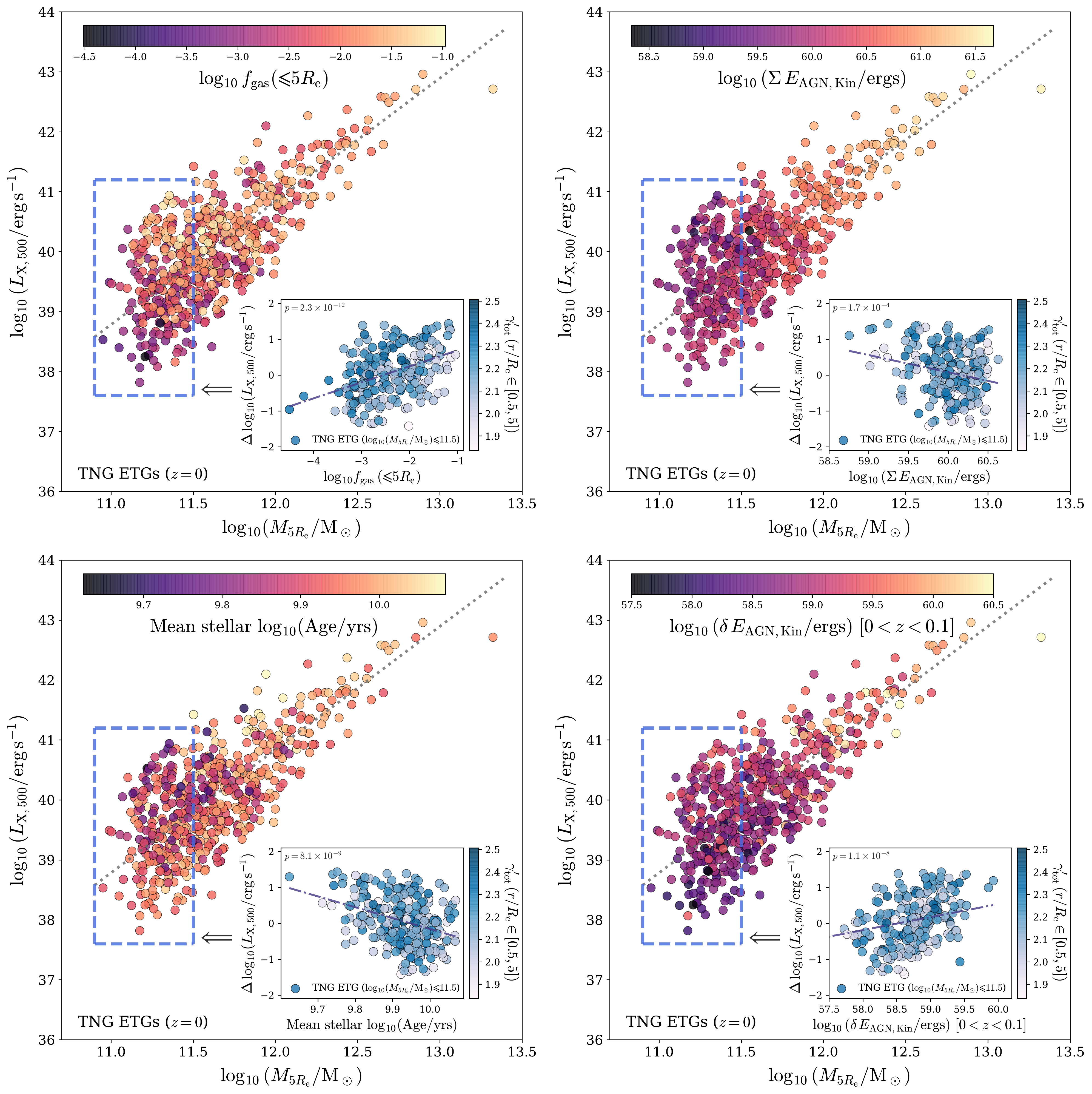}
    \caption{The $L_{\mathrm{X, 500}}$--$M_{5R_{\rm e}}$ scaling relation colored by the total gas mass fraction ($f_{\rm gas}$, {\it top left}), the cumulative kinetic mode AGN feedback energy from their central SMBHs ($\Sigma E_{\rm AGN, Kin}$, {\it top right}), the mean stellar age of stars ({\it bottom left}), and the kinetic mode AGN feedback energy in the past 1.3 Gyrs ($\delta E_{\rm AGN, Kin}[0<z<0.1]$, {\it bottom right}). In each panel, the grey dashed line denotes the linear fit to the $L_{\mathrm{X, 500}}$--$M_{5R_{\rm e}}$ scaling relation in the left panel of Fig.~\ref{fig:2}. The blue dashed boxes in each panel marks out the low-mass region ($\log_{10} (M_{5R_{\rm e}}/\mathrm{M_{\astrosun}}) \leqslant 11.5$) we focus on exploring the scatter in X-ray luminosity, and there are clear correlations between $L_{\rm X, 500}$ and the colored physical quantities in every box. In the {\it inset panels} for each subplot, we show the offset in the $L_{\mathrm{X, 500}}$ from the best-fit $L_{\mathrm{X, 500}}$--$M_{5R_{\rm e}}$ scaling relation, $\Delta L_{\mathrm{X, 500}}$, versus the {\it color bar} physical quantities for ETGs in the blue boxes. The blue dotted-dashed line in each inset plot is the linear fit to the low-mass ETGs with their Pearson $p$ values labelled at the top left corner of each inset. The color maps for the inset panels indicate the power-law slope of the total density profiles for these low-mass ETGs measured from $0.5R_{\mathrm{e}}$ to $5R_{\mathrm{e}}$. We truncate the $x$ axis lower limit at $57.5$ for the bottom right inset panel to focus on the main distribution of data points in our sample, but this truncation does hide 4 ETGs with $\log_{10} (\delta E_{\mathrm{AGN, Kin}}/\mathrm{ergs}) \ll 57.5$. The four {\it inset panels} further corroborate the significant correlations between $\Delta L_{\mathrm{X, 500}}$ and the four colored physical quantities in the main plots.}
    \label{fig:4}
\end{figure*}

In this section, we discuss the correlation between the offset from the best-fit $L_{\mathrm{X, 500}}$--$M_{5R_{\rm e}}$ scaling relation and the dark matter fraction within $5R_{\mathrm{e}}$ of the IllustrisTNG ETGs. The offset $\Delta L_{\mathrm{X, 500}}$ is calculated as the difference between the IllustrisTNG ETG data points and their linear fit (black line) in Fig.~\ref{fig:2}. This a natural way of quantifying the scatter off of the mean scaling relation following \citet{2019MNRAS.488.1072K}, while it also allows for characterizing secondary effects that affect $L_{\rm X}$ other than the primary factor, total mass.  \citet{2019MNRAS.488.1072K} found that for Chandra ETGs with $\log_{10} (M_{5R_{\rm e}}/\mathrm{M_{\astrosun}}) < 11.5$, those that up-scatter in $L_{\mathrm{X, 500}}$ at fixed total mass tend to have lower dark matter fraction (and vice versa for down-scattering $L_{\mathrm{X, 500}}$ ETGs, see their Figure 8 for details). This effect is stronger for cuspy Chandra ETGs (compared to cored ones) which are often fast rotators possessing younger stellar populations. They suggested that this difference for cuspy versus cored ETGs could be linked to differences in the strengths of stellar feedback in the two types of systems.

In the left panel of Fig.~\ref{fig:3}, we plot the $L_{\mathrm{X, 500}}$--$M_{5R_{\rm e}}$ scaling relation colored by $1-f_{\mathrm{star}} (\leqslant 5R_{\rm e})$ for IllustrisTNG ETGs. The blue dashed box indicates the galaxies with $\log_{10} (M_{5R_{\rm e}}/\mathrm{M_{\astrosun}}) \leqslant 11.5$, and at face value there is no significant vertical trend with $1-f_{\mathrm{star}}$ for galaxies above or below the linear fit. This is qualitatively different than in the \citet{2019MNRAS.488.1072K} sample (their Figure 8), however we do notice that their observed ETGs above the linear fit typically have $f_{\mathrm{DM}} < 0.5$, while the simulated ETGs have $f_{\mathrm{DM}} \gtrsim 0.6$. In Appendix~\ref{sec:App_A} we show that selecting `relaxed' and `unrelaxed' host haloes for our selected ETGs does not impact the range of their dark matter fractions, such that these high dark matter fractions are not artifacts of mis-centered ETGs residing in unrelaxed haloes.

In the right panel of Fig.~\ref{fig:3}, we plot the offset from the linear fit of the X-ray luminosity/total mass scaling relation, $\Delta L_{\mathrm{X, 500}}$, against the combined gas and dark matter mass fraction, $1-f_{\mathrm{star}} (\leqslant 5R_{\rm e})$, for {\it all} simulated ETGs. This definition follows \citet{2019MNRAS.488.1072K} where they assumed the gas mass fraction to be negligible compared to dark matter and defined $f_{\mathrm{DM}} = 1 - f_{\mathrm{star}}$ as a proxy for dark matter fraction. Our IllustrisTNG sample has gas mass fractions typically less than $1\%$ (median $0.8\%$) which agrees with their assumption (also see top left panel in Fig.~\ref{fig:4}). We also overplot the 61 observed ETGs from \citet{2019MNRAS.488.1072K} Figure 11 for comparison. The linear fits to the simulation and observation samples yield similar slopes and both show robust (very small Pearson $p$ values as labelled in the plots) negative correlations between $1-f_{\rm star}$ and $\Delta L_{\mathrm{X, 500}}$. At $1-f_{\rm star} \gtrsim 0.6$, the simulation also shows a similar level of scatter of $\Delta L_{\mathrm{X, 500}}$ (0.59 dex) as compared to observations (0.74 dex). Therefore, although IllustrisTNG reproduces the negative correlation between $\Delta L_{\mathrm{X, 500}}$ and the dark matter fraction, the compressed range of $1-f_{\rm star}$ leads to the lack of apparent correlation between $L_{\rm X, 500}$ and $1-f_{\rm star}$ in the top left panel of Fig.~\ref{fig:3}, especially for ETGs with $\log_{10} (M_{5R_{\rm e}}/\mathrm{M_{\astrosun}}) < 11.5$, where simulated ETGs both above and below the mean X-ray luminosity-mass relation can have similar $1-f_{\rm star}$.

Observationally-derived dark matter fractions are not free from systematics, and the range of dark matter fraction values can be dependent on the specific approach used for measuring the total mass of the ETGs. \citet{2018MNRAS.481.1950L} compared dark matter fractions within $5R_{\rm e}$ of IllustrisTNG galaxies to observations (top left panel of their Figure 12). IllustrisTNG showed better agreement with \citet{2013MNRAS.428.2407W} who used satellite galaxy kinematics compared to \citet{2017MNRAS.468.3949A} who used GC kinematics. Since most of the ETG masses in \citet{2019MNRAS.488.1072K} come from GC kinematics in \citet{2017MNRAS.468.3949A}, the tendency for ETGs in \citet{2019MNRAS.488.1072K} to have lower dark matter fractions than IllustrisTNG ETGs is consistent with expectations. Thus, if one adopted satellite kinematics-based dark matter fractions (mostly larger than $50\%$) from \citet{2013MNRAS.428.2407W} for the observed ETGs, the correlation between $\Delta L_{\mathrm{X, 500}}$ and $1-f_{\mathrm{star}}$ would likely weaken or even disappear.

As \citet{2019MNRAS.488.1072K} discovered that $\Delta L_{\rm X}$ varies differently with $1-f_{\rm star}$ for cuspy versus cored ETGs, we investigate the systematic trends of $\Delta L_{\mathrm{X, 500}}$ with the total mass density profile for our simulated ETGs.  Systematic covariance in dark matter fraction and total density profiles of ETGs is well-studied through observed stellar kinematics modeling~\citep{2007MNRAS.382..657T,2014MNRAS.445..115T,2017MNRAS.467.1397P,2018MNRAS.476.4543B,2021MNRAS.506.3691D}, strong gravitational lensing~\citep{2010ApJ...724..511A,2011ApJ...727...96R,2011MNRAS.415.2215B,2013ApJ...777...98S,2021MNRAS.503.2380S,2022arXiv220704070E}, and in hydrodynamic simulations~\citep{2017MNRAS.464.3742R,2017MNRAS.469.1824X,2020MNRAS.491.5188W}. The color map for the right panel of Fig.~\ref{fig:4} shows the power-law slope ($\gamma_{\rm tot}^{\prime}$) of the total density profile of {\it all} the IllustrisTNG ETGs measured in 100 logarithmic bins from $0.5R_{\mathrm{e}}$ to $5R_{\mathrm{e}}$. The color gradient in $\gamma_{\rm tot}^{\prime}$ is mainly visible along the $1-f_{\rm star}$ direction, while ETGs having different $\Delta L_{\mathrm{X, 500}}$ at fixed $1-f_{\rm star}$ do not seem to have drastically different density profiles. This indicates that the total density profile, at least down to $0.5 R_{\rm e}$ (limited by the simulation softening scale), does not play a dominant role in driving the scatter in $L_{\mathrm{X, 500}}$ of our IllustrisTNG ETG sample. Future higher resolution cosmological simulations that can resolve the inner density profiles for a large number of ETGs down to $0.1 R_{\rm e}$ would be desirable for further disentangling the X-ray gas properties of cuspy versus cored ETGs.

\subsection{The impact of AGN feedback}
\label{sec:3.3}

In this section, we discuss how variations in the cumulative kinetic (wind) mode AGN feedback energy for our IllustrisTNG ETGs drives the scatter in $L_{\rm X, 500}$ at the low-mass end, meanwhile also leading to covariances in gas fractions, stellar ages, and recent AGN activities for these low-mass ETGs. 

Supermassive blackholes (SMBH) at the centers of haloes can regulate gas properties including temperature and cooling timescales. Since the CGM (including the ISM within the central galaxy) is directly impacted by active galactic nucleus (AGN) feedback, we expect the effects of AGN feedback to also manifest in the scatter of $L_{\mathrm{X, 500}}$ for ETGs.  The IllustrisTNG AGN feedback model~\citep{2017MNRAS.465.3291W,2018MNRAS.479.4056W} features two channels: the radiative (pure thermal) mode that resembles high-accretion-rate thermal feedback common to high redshift quasars, and the low-accretion-rate kinetic mode that approximates centrally driven winds in quiescent AGN (no thermal energy injection). The former deposits thermal energy isotropically to the surrounding ISM of the central SMBH, while the latter injects kinetic kicks to the surrounding gas in random directions. For massive galaxies with $M_{\ast} \gtrsim 10^{11}\,\mathrm{M_{\astrosun}}$, the kinetic mode AGN feedback energy becomes the dominant feedback channel over the radiative mode at $z=0$~\citep{2020MNRAS.499..768Z}. \citet{2015MNRAS.449.4105C} also found that thermal AGN feedback alone cannot bring down $L_{\rm X}$ to observed values, and kinetic AGN feedback must be incorporated in order to further decrease $L_{\rm X}$ matching observed ETG X-ray scaling relations. Indeed, as we show in Appendix~\ref{sec:App_B}, this is also the case for our IllustrisTNG simulated ETG sample, such that the radiative AGN feedback energies do not significantly impact the scatter in $L_{\mathrm{X, 500}}$, and most of the SMBHs for our sample are thermally quiescent at $z<0.1$ (see middle and bottom panels of Fig.~\ref{fig:B}). Furthermore, through radiative heating, turbulence injection, kinetic expulsion of gas through winds, AGN feedback also lowers the star formation efficiency of gas in the central galaxy and becomes the dominant factor for quenching in massive central galaxies~\citep{2021MNRAS.500.4004D,2022MNRAS.512.1052P}.

In Fig.~\ref{fig:4}, we show how AGN feedback and its consequences on gas properties and star formation influences $L_{\mathrm{X, 500}}$, especially the scatter at $\log_{10} (M_{5R_{\rm e}}/\mathrm{M_{\astrosun}}) \leqslant 11.5$. The color maps include the gas mass fraction within $5 R_{\rm e}$ ($f_{\rm gas}$, top left), the mean stellar age (bottom left), cumulative feedback energy in the kinetic mode of the central SMBH ($\Sigma E_{\rm AGN, Kin}$, top right), and the kinetic mode feedback energy of the central SMBH from $z=0.1$ to $z=0$ ($\sim1.3$ Gyrs, $\delta E_{\rm AGN, Kin}$, bottom right). In the low-mass region with $\log_{10} (M_{5R_{\rm e}}/\mathrm{M_{\astrosun}}) \leqslant 11.5$ (blue dashed boxes, we refer to galaxies in this box as `low-mass ETGs' in the following), $L_{\rm X, 500}$ shows clear correlations with all four colored physical quantities. In the inset plots of each panel, we further show for ETGs in the blue boxes their offset from the mean scaling relation $\Delta L_{\rm X, 500}$ versus their respective colored quantities in the main plots. The linear fits and the very small Pearson $p$-values of the fits in the insets demonstrate robust  negative correlations between $L_{\rm X, 500}$ with mean stellar age and cumulative kinetic AGN feedback energy $\Sigma E_{\mathrm{AGN, Kin}}$, as well as robust positive correlations with the gas fraction $f_{\mathrm{gas}}$ and recent AGN kinetic feedback $\delta E_{\mathrm{AGN, Kin}} (0<z<0.1)$.

To self-consistently explain these four sets of systematic trends in the scatter of $L_{\rm X, 500}$ at the the low-mass end, we argue that the cumulative kinetic feedback energy from their SMBH is the driving factor. In this scenario, lower mass ETGs that up-scatter in $L_{\mathrm{X, 500}}$ experienced less AGN kinetic feedback historically and retained a larger gas reservoir, leading to higher gas mass fraction. This is consistent findings in previous work using IllustrisTNG that the X-ray luminosity positively correlates with the CGM fraction in the mass range $\log_{10} (M_{\ast}/\mathrm{M_{\astrosun}})\in [10.8, 11.4]$~\citep{2020ApJ...893L..24O}. The less-violent AGN (kinetic) feedback leads to younger stellar populations due to less efficient quenching. The consequence of having younger stars is stronger stellar feedback that can further heat up the ISM and CGM, leading to higher $L_{\rm X, 500}$. Moreover, the fact that these up-scatter ETGs in $L_{\mathrm{X, 500}}$ end up having larger $f_{\rm gas}$ also fuels more {\it recent} AGN accretion and hence {\it higher} recent feedback energies ($\delta E_{\rm AGN, Kin}$).  

Conversely, one could argue that stronger {\it recent} (instead of historic) AGN feedback or stellar feedback (star formation) is the dominant driver for creating the scatter in $L_{\rm X, 500}$ at the low-mass end. Since most of our ETGs have thermally quiescent AGN in the last 1.3 Gyrs (bottom panel of Fig.~\ref{fig:B}), stronger recent {\it kinetic} AGN feedback in up-scatter $L_{\rm X, 500}$ alone could not lead to hotter gas around these ETGs without effective radiation. If recent AGN kinetic feedback energy did have a significant impact on the ISM and CGM, the color gradients in the left two panels of Fig.~\ref{fig:4} should revert, with up-scatter $L_{\rm X, 500}$ ETGs having lower $f_{\rm gas}$ and old stellar ages. Similarly, if stronger stellar feedback is the core driver of the scatter in $L_{\rm X, 500}$ at the low-mass end, it should also drive stronger outflows and lead to lower $f_{\rm gas}$ in up-scatter $L_{\rm X, 500}$ ETGs, which is opposite from the top left panel in Fig.~\ref{fig:4}. It will also be hard to explain why systematic variations in star formation or stellar feedback also simultaneously cause variations in the AGN feedback energies.  

Furthermore, we check in Appendix~\ref{sec:App_B} that $L_{\rm X, 500}$ and the SMBH mass $M_{\rm BH}$ do not have significant correlation for the low-mass ETGs with $\log_{10} (M_{5R_{\rm e}}/\mathrm{M_{\astrosun}}) < 11.5$. Thus, the variations in cumulative and recent AGN kinetic feedback energies in Fig.~\ref{fig:4} are not driven by variations in SMBH mass, but rather the diverse accretion and merger histories of the SMBH leading to different feedback histories that shape the scatter in $L_{\mathrm{X, 500}}$. Our findings here are also consistent with the previous study of \cite{2021MNRAS.501.2210T}, who showed that IllustrisTNG galaxies having $L_{\mathrm{X, 500}} \lesssim 10^{41} \mathrm{erg\,s^{-1}}$ have significantly larger scatter in $M_{\rm BH}$ at fixed $L_{\rm X}$ than galaxies with $L_{\mathrm{X, 500}} \gtrsim 10^{41} \mathrm{erg\,s^{-1}}$, leading to a weaker correlation between $L_{\rm X}$ and $M_{\rm BH}$.

Although many observations suggest that the internal heating of X-ray gas in ETGs is driven by stellar feedback at low masses~\citep{2006ApJ...653..207D,2019MNRAS.488.1072K,2020MNRAS.492.2095K}, our findings above dive deeper and provide a more comprehensive picture about the fundamental physics behind the scatter in $L_{\rm X}$.  Weaker kinetic AGN feedback in the past could enhance star formation and gas fractions, leading to present-day younger stellar populations and as a consequence, drive stronger stellar feedback that heat up the gas. However, this is qualitatively different from stellar feedback being the intrinsic source of feedback that leads to higher $L_{\rm X}$. Since observations are limited to static AGN properties, they may seem obvious when compared to observables (e.g. star formation rate, stellar ages, metallicity) that probe stellar feedback. We also point out that \citet{2011ApJ...729...12B} observed a slightly positive correlation of $L_{\rm X}$ with stellar age at around $K$-band magnitude of $\sim 10^{11}\,\mathrm{L_{\astrosun}}$, opposite from what we see in IllustrisTNG. Since they had only 7 galaxies, their result could suffer from small number statistics. But if future observations with better statistics still see older stellar populations in ETGs with larger $L_{\rm X}$ at fixed masses, it may suggest limitations in the current IllustrisTNG AGN model and advocate for more efficient radiative feedback at low redshift that can simultaneously quench star formation (older ages) and heat up the gas (higher $L_{\rm X}$).

Finally, the color maps in the inset plots in Fig.~\ref{fig:4} mark out the variations of the total density profile for the low-mass ETGs. There is no significant correlation between $\Delta L_{\mathrm{X, 500}}$ with the total density profile power-law slope. Reflecting on \citet{2019MNRAS.490.5722W}, the evolution of the total density profile in massive ETGs are mainly dominated by gas-poor mergers at $z\lesssim 1$. AGN feedback only impacts the total density profile at $z \gtrsim 1$ when the ETGs were still quite gaseous, while the efficient kinetic mode feedback can impact the small fraction of diffuse gas all the way to $z=0$, shaping $\Delta L_{\mathrm{X, 500}}$. These different physical origins of $\Delta L_{\mathrm{X, 500}}$ and $\gamma_{\rm tot}^{\prime}$ at low redshift make their dis-correlation a natural outcome.

\section{Outliers of the $L_{\mathrm{X}}$--$T_{\mathrm{sl}}$ relation: probing backsplash}
\label{sec:4}

\begin{figure}
	\includegraphics[width=\columnwidth]{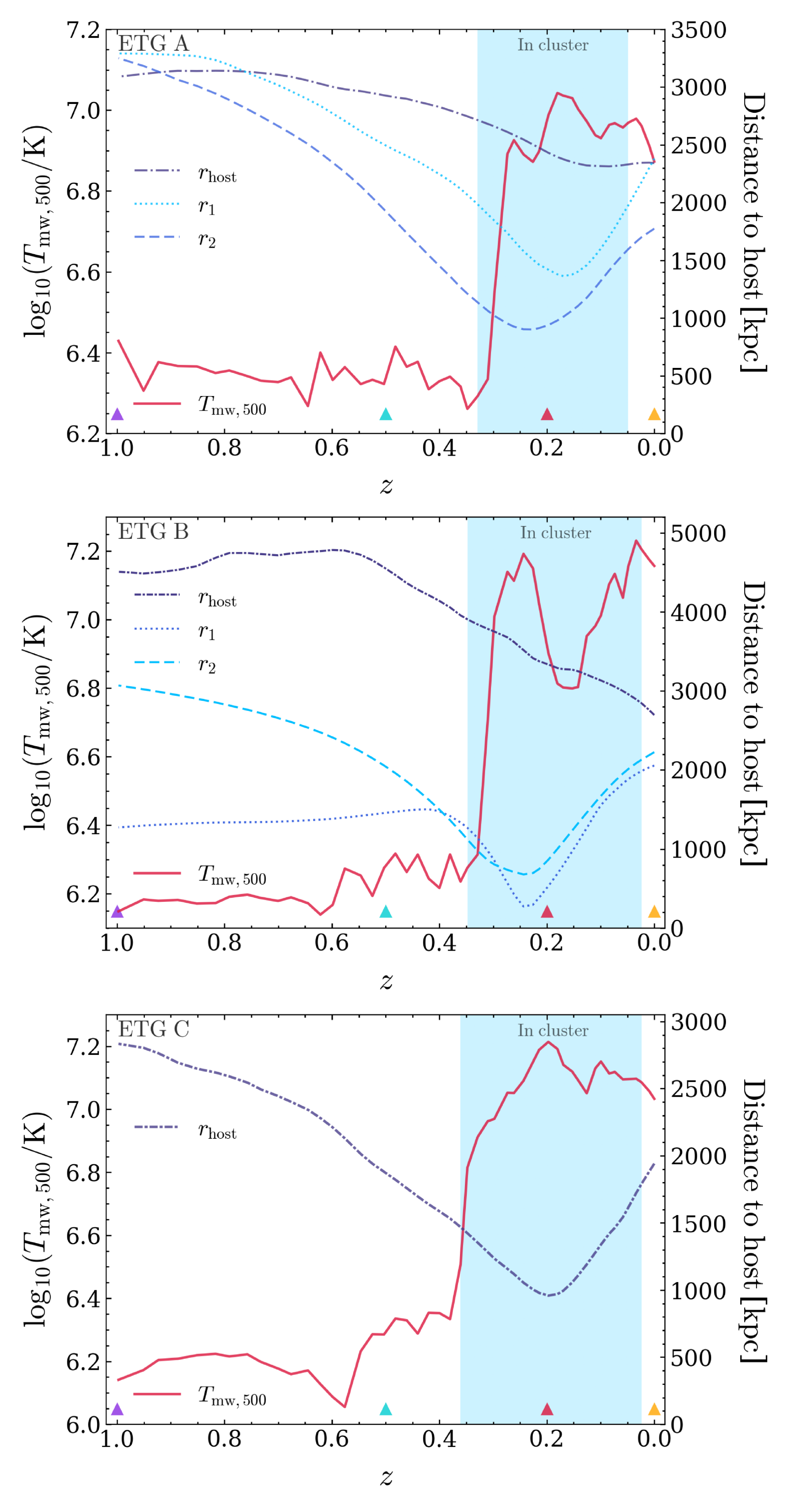}
    \caption{This figure shows the mass-weighted temperature history and distances to host haloes they interacted with for the three outlier ETGs in Fig.~\ref{fig:2}. In each panel, the mass-weighted temperature tracks are denoted by the solid red curve and refers to the left $y$-axis ($T_{\mathrm{mw, 500}}$). The distances to the host FoF groups ($r_{\mathrm{host}}$) and other major subhaloes ($r_{1}$, $r_{2}$ for ETGs A and B) they interacted with are indicated with the blue curves that refer to the right $y$-axis. The blue shaded region in each panel indicates the period when each ETG was inside the larger cluster they interacted with. The colored triangles stand for the four redshifts $z=0, 0.2, 0.5, 1$ at which we show temperature maps in Fig.~\ref{fig:6} and gas temperature and density profiles in Fig.~\ref{fig:7}.}
    \label{fig:5}
\end{figure}

In this section we investigate the formation history of the three outlier ETGs of the $L_{\mathrm{X, 500}}$--$T_{\mathrm{sl, 500}}$ relation as indicated in the right panel of Fig.~\ref{fig:2}. We conjecture that these objects are backsplash galaxies~\citep{2014ApJ...789....1D,2014JCAP...11..019A,2015ApJ...810...36M,2016ApJ...825...39M,2017ApJ...841...34M,2021MNRAS.504.4649O,2023MNRAS.520..649B} now in the field that were environmentally heated by their interactions with massive groups or clusters. We trace the distance of these three ETGs to all the more massive FoF groups it historically belonged to and how their temperatures evolved in the same period ($0<z<1$) along their main progenitor branches of their merger trees. For tracing gas temperature, we adopt the mass-weighted temperature definition following \citet{1996MNRAS.283..431B,2001ApJ...546..100M}:
\begin{equation}
    \label{eq:3}
    T_{\mathrm{mw}} = \frac{\sum_{i}m_{i} T_i}{\sum_{i}m_{i}}\,,
\end{equation}
where $m_{i}$ and $T_{i}$ are the mass and temperature of the $i$-th gas cell. The reason why we choose $T_{\mathrm{mw}}$ over $T_{\mathrm{sl}}$ for temperature tracing is that, in the event of shock heating (from falling into a more massive halo), the heated cells lose weight rapidly due to the $T^{-3/4}$ scaling in the $T_{\mathrm{sl}}$ definition. This dials down the effect of heating as measured by $T_{\mathrm{sl}}$, while the mass-weighted temperature $T_{\mathrm{mw}}$ is not sensitive to the gas cell temperature. There is also better physical motivation to trace the mass-weighted temperature, as the total internal energy of the gas $U\propto mT_{\mathrm{mw}}$ and reflects the energetic state of the gas in the CGM. Similar to the $T_{\rm sl}$ definition in Section~\ref{sec:2.3}, we also neglect the very cold and dense gas cells with temperature $T_{i}<0.05\,\mathrm{keV}$ ($\sim 5.8\times 10^{5}\,\mathrm{K}$), number density $n_{\mathrm{H}, i}>0.1$, or star formation rate SFR$>0$ when calculating $T_{\rm mw]}$. We show a comparison in Appendix~\ref{sec:App_C} between the spectroscopic-like temperature and mass-weighted temperature for all 559 ETGs in our sample. The $T_{\mathrm{mw, 500}}$ definition yields $\sim 0.2$ dex higher temperatures than the $T_{\mathrm{sl, 500}}$ definition for our ETGs.

\subsection{Temperature history and interactions with larger clusters}
\label{sec:4.1}

\begin{figure*}
	\includegraphics[width=1.75\columnwidth]{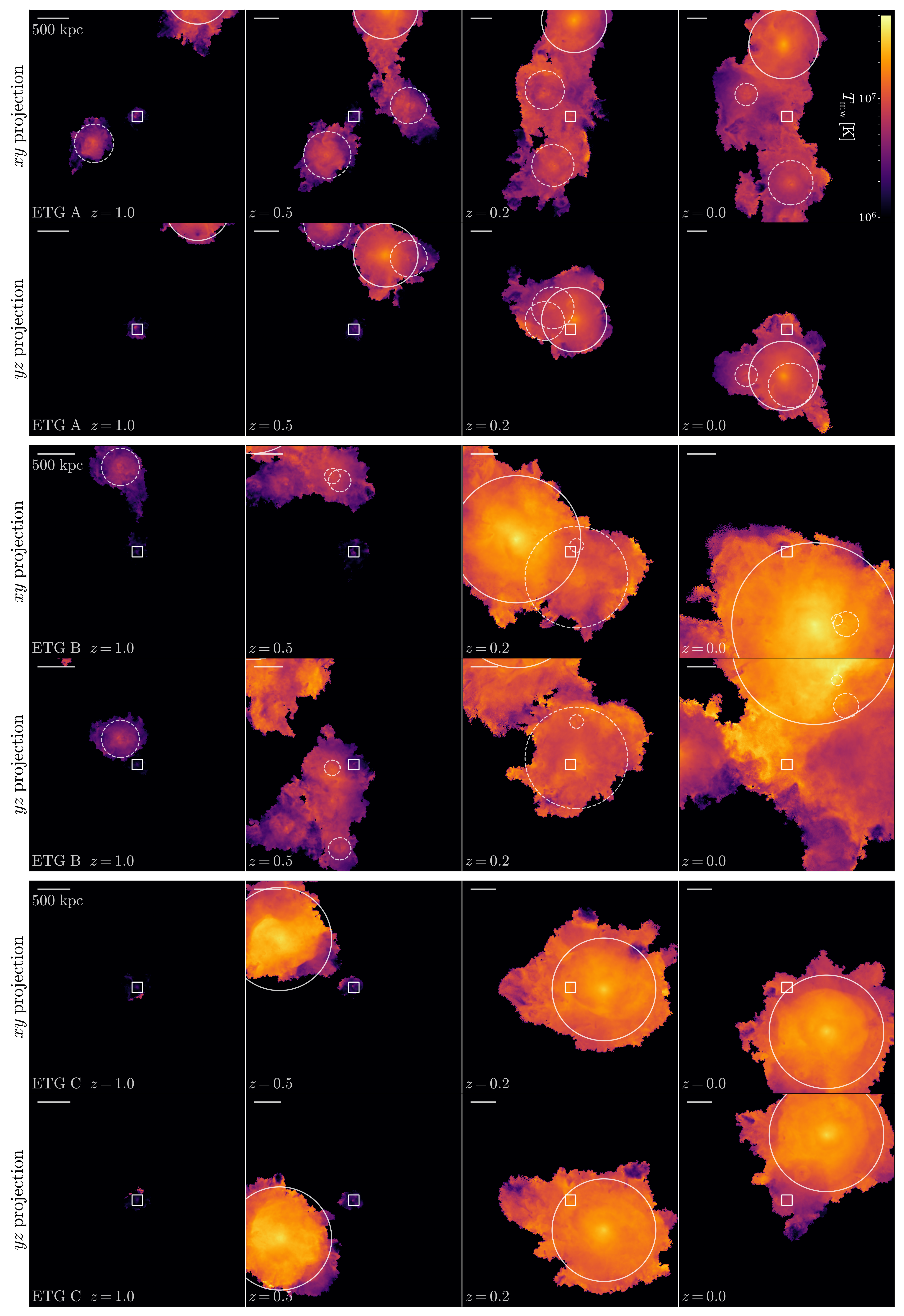}
    \caption{The projected mass-weighted temperature maps  at $z=1, 0.5, 0.2, 0$ for ETG A ({\it top panel}), B ({\it middle panel}), and C ({\it bottom panel}) and the large clusters they interacted with. In each panel, the top row shows the gas temperature in the $x$--$y$ plane while the bottom row in the $y$--$z$ plane, with both rows taking the $y$-axis in the horizontal direction. ETGs A, B, and C in each subplot are fixed in the center (white squares); the cluster host halo $R_{200}$ are marked by solid white circles; the $2\times$ stellar-half-mass radius of the subhalos with which these three ETGs interacted with are marked by dashed white circles. The scale bars in the top left corner of each temperature maps stand for 500 physical kpc. The color map for all subplots range from $10^6$ K to $10^{7.7}$ K as shown in the top right corner. These gas temperature maps further elucidate the context of environmental heating for the three outlier ETGs.}
    \label{fig:6}
\end{figure*}

\begin{figure*}
	\includegraphics[width=2\columnwidth]{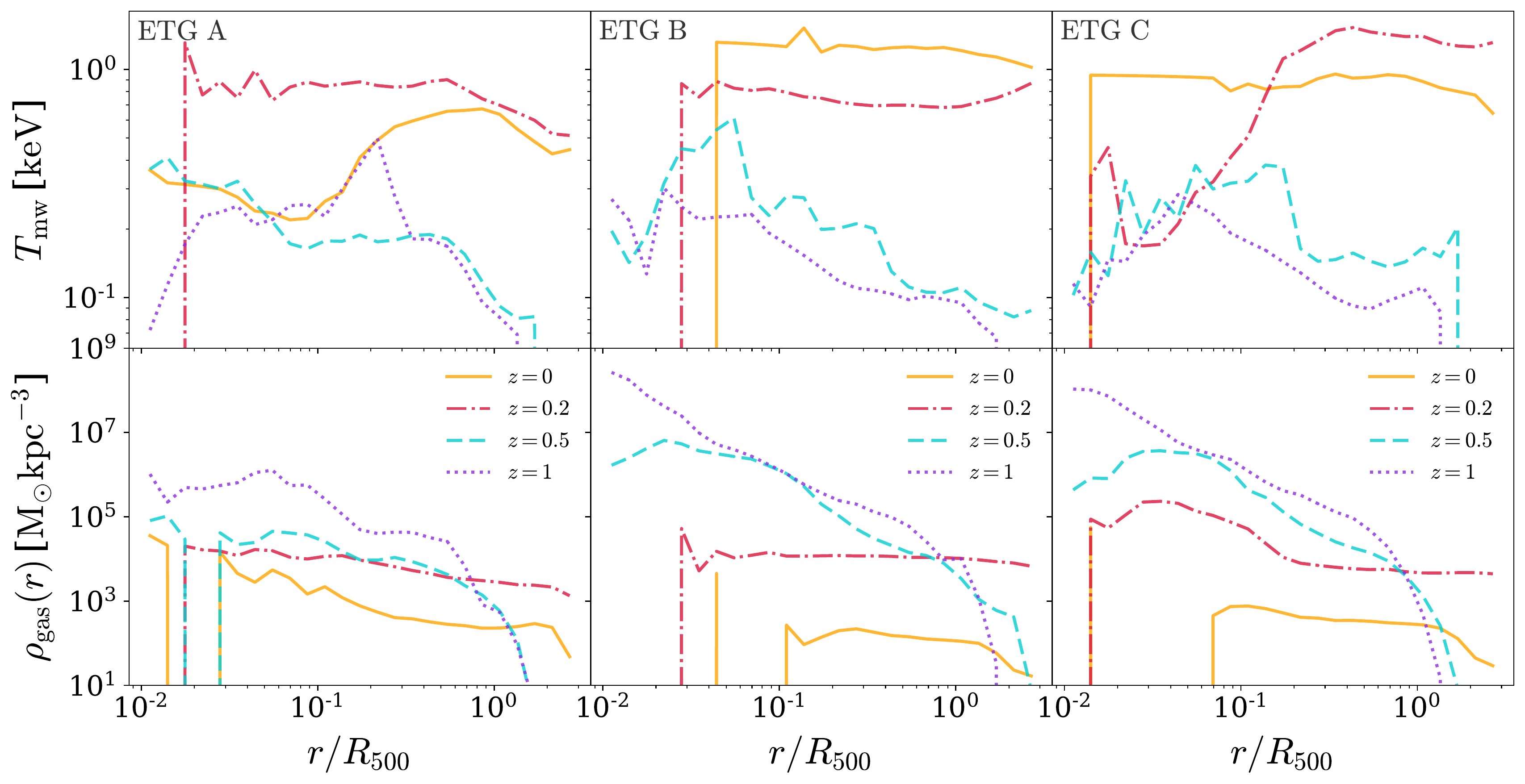}
    \caption{The gas mass-weighted temperature (top row) and gas density (bottom row)  profiles within $3\times R_{500}$ for the three outlier ETGs we studied in Fig.~\ref{fig:6}, with each column showing one galaxy. The $R_{500}$ values are for the host haloes of the three ETGs at $z=0$, $z=0.5$, and $z=1$ when they were the central galaxies. We take the average value of the pre-infall and post-infall $R_{500}$ of the ETG as their approximate $R_{500}$ at  $z=0.2$. The four sets of curves in each row indicate the gas temperature and density profiles at four different redshifts, i.e. $z=0, 0.2, 0.5, 1$. These four redshifts correspond to the epochs shown in Fig.~\ref{fig:6}, which cover before entering, right within, and after exiting the more massive clusters each ETG interacted with. Through these interactions, the temperature profiles develop very hot outskirts and the gas density becomes very sparse.}
    \label{fig:7}
\end{figure*}

In Fig.~\ref{fig:5}, we show the $T_{\mathrm{mw, 500}}$ evolution history for the three outlier ETGs from $z=1$ to $z=0$. Their gas mass-weighted temperatures all seem to jump up at $z\sim0.35$. By examining their merger trees and cluster (FoF group) memberships, we find that they all belonged to a much larger cluster during the periods indicated by the vertical dashed lines in each panel. They all interacted with larger galaxy groups/clusters since $z\sim 0.35$, lost their central ETG identity during their infall, and came back to become backsplash centrals just before $z=0$. 

We also show their distances to the host FoF group and major subhaloes of those clusters (subhaloes by $z=0$) that they interacted with as a function of time (blue curves, right $y$-axis in Fig.\ref{fig:5}). The temperature rise of each ETG coincides well with their first accretion into a larger cluster, and temperature peaks correspond to the closest encounters with the brightest cluster galaxy (BCG) or other member galaxies, indicative of environmental heating of their gas. To further demonstrate this process, we plot the mass-weighted temperature maps at $z=1, 0.5, 0.2, 0$ in the surroundings of each ETG in Fig.~\ref{fig:6}. We only include gas cells associated with the ETGs and the larger cluster they interacted with in these temperature maps. We analyze how the three ETG formation histories impact the heating of their gas on a case-by-case basis:
\begin{itemize}
    \item ETG A (\textsc{Subfind} ID 410700) interacted with parts of a FoF group that has a $z=0$ $\log_{10} (M_{200}/\mathrm{M_{\astrosun}}) = 13.95$. Specifically, it mainly interacted with two subhaloes of the host FoF group, the first one ($r_1$) having a $z=0$ total mass of $\log_{10} (M_{\rm sub}/\mathrm{M_{\astrosun}}) = 13.44$, and the second one ($r_2$) having a $z=0$ total mass of $\log_{10} (M_{\rm sub}/\mathrm{M_{\astrosun}}) = 13.02$. The two temperature peaks of the ETG at $z\sim0.25$ and $z\sim0.2$ coincide well with the pericentric passages with these two subhaloes haloes, after which the gas in the ETG remained $\sim 0.7$ dex hotter than pre-infall. From the temperature maps we can see that the ETG (white squares) started interacting with outskirt gas in subhalo $r_2$ at $z=0.5$, heated up and remained to be surrounded by extended hot gas from the host halo of the larger FoF group even after exiting at $z=0$.

    \item ETG B (\textsc{Subfind} ID 482814) interacted with the largest cluster in TNG100, which has a $z=0$ $\log_{10} (M_{200}/\mathrm{M_{\astrosun}}) = 14.58$. It first interacted with two subhaloes of the cluster, the first one ($r_1$) having a $z=0$ total mass of $\log_{10} (M_{\rm sub}/\mathrm{M_{\astrosun}}) = 13.06$, and the second one ($r_2$) having a $z=0$ total mass of $\log_{10} (M_{\rm sub}/\mathrm{M_{\astrosun}}) = 12.48$. The ETG hit pericenter with these two subhaloes almost at the same time of $z\sim 0.25$, corresponding to the first temperature peak after which the gas starts to cool. The second peak towards $z=0$ in the temperature history comes from the final approach of the ETG onto the most massive cluster, when the other two subhaloes already finished their infall. Although the ETG is a central galaxy of a individual FoF group at $z=0$, it is in a multi-merger series happening in the densest region of the box and is impacted by the extremely hot and extended gas of the largest cluster in the simulation. 

    \item ETG C (\textsc{Subfind} ID 486341) interacted directly with the host halo of a FoF group that has a $z=0$ $\log_{10} (M_{200}/\mathrm{M_{\astrosun}}) = 14.32$. Again, the gas temperature in the ETG starts to rise rapidly after becoming part of the larger FoF group, and its temperature peak at $z\sim 0.2$ corresponds well to the pericenter passage with the host halo. After exiting the FoF group at $z=0.02$, the ETG still submerges in the extended hot gas of the cluster and maintains $\sim 0.8$ dex higher gas temperature than its pre-interaction temperature. 
\end{itemize}

These three case studies clearly demonstrate that the outliers in the $L_{\rm X}$--$T_{\rm sl}$ relation originate from environmental heating of the hot gas through interactions with larger galaxy clusters. Since they are backsplash ETGs that are relatively low in mass, their change in potential energy are insufficient to cause the order-of-magnitude increase in temperature due to gravitational heating. The presence of a virial shock~\citep{2003MNRAS.345..349B} in the host halo can heat the accreted gas to the virial temperature of the host halo. Although some of the encounters above do not penetrate deep into the halo, we find that the extended gas outside of $R_{200}$ of the larger clusters these ETGs interacted can still have much higher temperatures that supports external heating~\citep{2020MNRAS.492.2095K}. Although observers~\citep{2016ApJ...826..167G} did not find significant temperature differences for ETGs in cluster/group versus field environments, indicating that environmental heating may not impact the `main ridge' of the $L_{\rm X}$--$T_{\rm sl}$ relation, shock heating and ram-pressure compression can still work in special circumstances as in our three outlier ETGs that heat up the remaining gas during merger interactions.

For massive $M_{\ast} > 10^{10}\,\mathrm{M_{\astrosun}}$ galaxies experiencing tidal interactions in IllustrisTNG, \citet{2022arXiv220907711L} found that the ISM ($r\in[0, 2R_{\rm e}]$) and CGM ($r\in[2R_{\rm e}, 4R_{\rm e}]$) can get stripped by $\gtrsim 70\%$ in the absence of shocks and $\gtrsim 90\%$ with shocks, leading to significant removal of the total gas content. Indeed, in Fig.~\ref{fig:2} left panel, the $L_{\rm X}$ of these three outlier ETGs is all below the best-fit $L_{\mathrm{X, 500}}$--$M_{5R_{\rm e}}$ relation, indicating gas stripping in all three cases. Therefore, the combination of environmental heating raising $T_{\rm sl}$ and tidal/ram-pressure stripping lowering $L_{\rm X}$ (removing gas content) during these mergers/interactions with larger clusters lead to these significant outliers of the $L_{\rm X}$--$T_{\rm sl}$ scaling relations. 

Through visually examining the gas temperature and FoF group membership history of each ETG in our sample, we found {\it four} other ETGs that similarly had encounters with more massive FoF groups but lie on the main sequence of the $L_{\rm X}$--$T_{\rm sl}$ scaling relation. We find that three of these four `normal' ETGs had very large $M_{500}$ mass ratios ($>1/2$) with their host halos and their gas temperatures only increased $\sim 0.3$ dex during these interactions. The remaining `normal' ETG interacted with a massive cluster with $M_{500}\sim 10^{13.7}\,\mathrm{M_{\astrosun}}$ towards $z=0$, which should have made it an outlier ETG in the $L_{\rm X}$--$T_{\rm sl}$ plane as with ETGs A, B, and C. However, this ETG was only part of the larger FoF group for $\lesssim 1$ Gyr and brushed the cluster outskirts ($>2.8$ Mpc), rendering the environmental heating from the host ineffective. Therefore, backsplash ETGs need to have close interactions with massive clusters ($M_{500}\sim 10^{14}\,\mathrm{M_{\astrosun}}$) in order to be detectable as $L_{\rm X}$--$T_{\rm sl}$ scaling relation outliers.

\subsection{Gas temperature and density profiles}
\label{sec:4.2}

To elaborate on the scenario described above, we further show in Fig.~\ref{fig:7} the mass-weighted temperature and density profiles of gas for the three outlier ETGs at $z=0, 0.2, 0.5, 1$. In these profiles, we use all gas particles that are enclosed within $3R_{500}$ of the ETG host halos at that redshift (for $z=0.2$ when these ETGs where in larger FoF groups and their own $R_{500}$ could not be defined, we use the average value of their pre-infall and post-infall $R_{500}$ values). The triangles in Fig.~\ref{fig:6} mark these epochs and correspond to times before infall, while within, and just after exiting the more massive FoF groups each of these ETGs interacted with. 

At $z=1$ and $z=0.5$ before entering the larger clusters they interacted with, ETG A has an overall flat mass-weighted temperature profile within $R_{500}$, while ETGs B and C have decreasing temperature profiles with increasing radius. The density profiles show that all three ETGs started out with cuspy gas profiles at $z=1$ (especially ETG B and C) within $r\sim 0.1 R_{500}$ and the outer density profiles follow steep power-law-shaped profiles.

At $z=0.2$ when these three ETGs were within larger FoF groups, their temperature profiles are significantly boosted. For ETG A, the temperature profile is almost flat and everywhere hotter than before infall. Most of the gas in ETG B also has a hot and flat temperature profile at $z=0.2$, and even a small rise in temperature beyond $R_{500}$. This is a clear signature for the onset of external heating~\citep{2020MNRAS.492.2095K} due to the interactions. External heating is again most evident in ETG C, where the gas within $r\lesssim 0.03 R_{500}$ still has similar temperatures with pre-infall ($z=0.5$), while the gas temperature signifcantly rises from $0.03 R_{500}$ to $0.2 R_{500}$, beyond which the gas temperature is much hotter and flat with radius. The density profiles show unanimous boosting of the outer gas density in all three ETGs at $z=0.2$ at $r \gtrsim 0.5 R_{500}$. Comparing the spatial extent of the hot gas of the larger FoF groups these ETGs interacted with in Fig.~\ref{fig:6}, the much hotter gas in the larger clusters is almost a constant background medium to these ETGs during the interactions. The gas density in the core region ($r\lesssim 0.1 R_{500}$) for all three ETGs decrease as compared to $z=0.5$, indicative of external (shock) heating and gas stripping.

Finally at $z=0$, when these three ETGs splashed-back from the larger FoF groups and became BCGs of their own groups, they all show significantly sparser gas density profiles compared to that before infall or during the interactions. Cooling in the center of ETG A leads to a flat cooler core and rising temperature at the outskirts, preserving signatures of environmental heating. The temperature profiles for ETG B and C both are hot and flat at $z=0$. The former (ETG B) has a final temperature higher than the $z=0.2$ temperature profile as it is starting its final approach onto the largest cluster in the TNG100 box at $z=0$ and keeps heating up (also see Fig.~\ref{fig:6}, middle row). The outskirts of the latter (ETG C) has a cooler temperature at $z=0$ than during the interaction at $z=0.2$ as the gas starts to cool after exiting the large cluster. 

These gas temperature and density profiles further support the environmental heating scenario described in Section~\ref{sec:4.1} as the cause of them being significant outliers in the $L_{\rm X}$--$T_{\rm sl}$ scaling relation. We argue that similar outliers seen in observed ETG X-ray scaling relations are also backsplash objects that might have recently had their gas heated up environmentally. This finding can also potentially facilitate systematic searches of backsplash objects around the outskirts of large galaxy clusters using next generation X-ray telescopes (e.g., eROSITA~\citealt{2012arXiv1209.3114M}, Athena+~\citealt{2013arXiv1306.2307N}, Lynx~\citealt{2019JATIS...5b1001G} etc.) by looking for outliers in the $L_{\rm X}$--$T_{\rm sl}$ scaling relation. The three outlier ETGs we investigated all fell into massive clusters at $z\sim 0.35$ and became backsplash galaxies at $z=0$. This coincides well with the $2\times$ dynamical timescale for dark matter halos at $z=0$ which is $3.89$ Gyrs ($z=0.37$). Therefore, we expect these environmentally heated ETGs to stand out as outliers on the $L_{\rm X}$--$T_{\rm sl}$ relation for approximately two dynamical timescales at their respective redshifts. Future work combining redshift-dependent halo merger rates with subhalo pericenter distance distributions could yield quantitative forecasts for the detection rate of backsplash ETGs in upcoming X-ray surveys.

\section{Conclusions}
\label{sec:5}

In this paper, we have explored the X-ray scaling relations of hot gas in massive ETGs from IllustrisTNG-100. We derived mock X-ray luminosities (using the public code MOCK-X) and spectroscopic-like temperatures for a legacy mock-ETG sample from \citet{2020MNRAS.491.5188W} that has well-studied total density profiles and dark matter fractions. We compared the X-ray luminosity-mass relation and the X-ray luminosity-temperature relation to observations for the simulated ETGs. We further studied how the low-mass end scatter in the X-ray luminosity-mass scaling relation relates to the ETG dark matter fraction, gas fraction, stellar population, and AGN feedback activities. Our major findings are as follows:

\begin{enumerate}
    \item The $L_{\mathrm{X, 500}}$--$M_{5R_{\rm e}}$ scaling relation has a very similar slope and overall scatter to the observed Chandra ETGs. The scatter increases at the low-mass end compared to the high-mass end. The $L_{\mathrm{X, 500}}$--$T_{\mathrm{sl, 500}}$ scaling relation has slightly steeper slope than the observed Chandra ETGs, though hot gas in IllustrisTNG ETGs are on average $\sim 0.5$ dex cooler than observations (Fig.~\ref{fig:2}).
    
    \item We do not observe significant correlation between $L_{\mathrm{X, 500}}$ and the ETG dark matter fraction (including the small amount of gas) $1-f_{\rm star}$ at the low-mass end (Fig.~\ref{fig:3} left). The offset from the best-fit X-ray luminosity-mass relation $\Delta L_{\mathrm{X, 500}}$ negatively correlates with $1-f_{\rm star}$ similar to observations (Fig.~\ref{fig:3} right), although IllustrisTNG ETGs have systematically larger dark matter fractions than observations, which was already known in earlier literature.
    
    \item At the low-mass end ($\log_{10} (M_{5R_{\rm e}}/\mathrm{M_{\astrosun}}) \leqslant 11.5$) ETGs that up-scatter in $L_{\mathrm{X, 500}}$ (vice versa the following for down-scatter) tend to have higher gas fractions and younger stellar populations. This is caused by their AGN having lower cumulative kinetic feedback energy, which leads to their higher gas content and less efficient quenching. As a consequence, the younger stars could also provide stronger stellar feedback heating that further increases $L_{\rm X}$. The higher gas fraction in the end fuels more active recent black hole accretion, leading to these up-scatter ETGs having stronger recent ($z<0.1$) AGN kinetic feedback (Fig~\ref{fig:4}). 
    
    \item Past AGN feedback mediates star formation and stellar populations in ETGs, thereby influencing the strength of stellar feedback at present-day. This can partially explain why observations see a sub-dominant role of AGN compared to stellar feedback with only access to $z=0$ AGN properties and not their full accretion history (Section~\ref{sec:3.3}).
    
    \item The scatter of $L_{\mathrm{X, 500}}$ does not correlate with the total density profiles of the ETGs at the low mass end. This indicates that mergers, which dominate the ETG density profile evolution at $z\lesssim 1$, do not significantly impact $L_{\mathrm{X, 500}}$ at fixed $M_{5R_{\rm e}}$ (Figs.~\ref{fig:3}, \ref{fig:4}).

\end{enumerate}

We also investigated the merger histories of three individual ETGs  (Fig.~\ref{fig:2} black squares) that are outliers of the X-ray luminosity-temperature scaling relation. We find that:
\begin{enumerate}
    \item All three ETGs interacted with much larger galaxy groups/clusters since $z\sim 0.4$ and were environmentally heated. Their gas temperature rise coherently with their infall into these larger FoF groups, and the temperature peaks coincide well with pericenter passages with those larger galaxy clusters (Fig.~\ref{fig:6}). Their gas temperature and density profiles also show clear signatures of external heating, leaving those ETGs with very sparse and hot gas atypical of their mass at $z=0$ (Fig.~\ref{fig:7}).
    
    \item Given that the three outlier ETGs are backsplash galaxies, we propose that querying outliers of the $L_{\rm X}$--$T_{\rm sl}$ relation can potentially be used as a search strategy for backsplash ETGs near galaxy clusters.
    
\end{enumerate}

As a final note, AGN feedback impacting the scatter in the low-mass-end of the  $L_{\mathrm{X, 500}}$--$M_{5R_{\rm e}}$ relation can be regarded as a theoretical prediction that could be verified in future observations correlating X-ray gas to the AGN of ETGs. New observations will provide important constraints for developing more realistic AGN feedback models in hydrodynamic simulations (e.g. more measurements for the stellar ages of low-mass X-ray ETGs) and testing the picture we have developed.. As for finding backsplash objects using outliers in the $L_{\rm X}$--$T_{\rm sl}$ relation, future X-ray missions can target the outskirts of galaxy clusters for red galaxies (cross-correlating with optical surveys) that simultaneously possess a hot X-ray component. Care should be taken as the signal-to-noise of such searches maybe limited to the specific instrument and the success of the proposed backsplash search campaign depends on the sensitivity to the contrast between the backsplash ETG gas and the extend hot gas from the large galaxy cluster. Nonetheless, our work here still provides a promising novel approach to finding backsplash galaxies in addition to the existing density profile-based~\citep{2021ApJ...923...37A} or surface-brightness-fluctuation-based~\citep{2022arXiv221100629C} inference methods that could yield fruitful outcomes in the near future.

\section*{Acknowledgements}

We would like to acknowledge the past contributions of David Barnes to this paper before a career shift beyond astronomy. We thank Ethan Nadler, Hui Li, Phil Mansfield, Risa Wechsler, and Shy Genel for helpful discussions and comments while preparing this draft. 
YW acknowledges the support of a Stanford-KIPAC Chabolla Fellowship. MV acknowledges support through NASA ATP 19-ATP19-0019, 19-ATP19-0020, 19-ATP19-0167, and NSF grants AST-1814053, AST-1814259, AST-1909831, AST-2007355 and AST-2107724. This research made use of computational resources at the MIT and Harvard research computing facilities supported by FAS and MIT MKI; the authors are thankful for the support from the FAS Research Computing and MIT Engaging technical teams. The flagship simulations of the IllustrisTNG project used in this work have been run on the HazelHen Cray XC40-system at the High Performance Computing Center Stuttgart as part of project GCS-ILLU of the Gauss Centre for Supercomputing (GCS). This research made extensive use of \href{https://arXiv.org}{arXiv.org} and NASA's Astrophysics Data System for bibliographic information.




\bibliographystyle{mnras}
\bibliography{xray} 

\begin{thebibliography}{}
\makeatletter
\relax
\def\mn@urlcharsother{\let\do\@makeother \do\$\do\&\do\#\do\^\do\_\do\%\do\~}
\def\mn@doi{\begingroup\mn@urlcharsother \@ifnextchar [ {\mn@doi@}
  {\mn@doi@[]}}
\def\mn@doi@[#1]#2{\def\@tempa{#1}\ifx\@tempa\@empty \href
  {http://dx.doi.org/#2} {doi:#2}\else \href {http://dx.doi.org/#2} {#1}\fi
  \endgroup}
\def\mn@eprint#1#2{\mn@eprint@#1:#2::\@nil}
\def\mn@eprint@arXiv#1{\href {http://arxiv.org/abs/#1} {{\tt arXiv:#1}}}
\def\mn@eprint@dblp#1{\href {http://dblp.uni-trier.de/rec/bibtex/#1.xml}
  {dblp:#1}}
\def\mn@eprint@#1:#2:#3:#4\@nil{\def\@tempa {#1}\def\@tempb {#2}\def\@tempc
  {#3}\ifx \@tempc \@empty \let \@tempc \@tempb \let \@tempb \@tempa \fi \ifx
  \@tempb \@empty \def\@tempb {arXiv}\fi \@ifundefined
  {mn@eprint@\@tempb}{\@tempb:\@tempc}{\expandafter \expandafter \csname
  mn@eprint@\@tempb\endcsname \expandafter{\@tempc}}}

\bibitem[\protect\citeauthoryear{{Abbott} et~al.,}{{Abbott}
  et~al.}{2018}]{2018PhRvD..98d3526A}
{Abbott} T.~M.~C.,  et~al., 2018, \mn@doi [\prd] {10.1103/PhysRevD.98.043526},
  \href {https://ui.adsabs.harvard.edu/abs/2018PhRvD..98d3526A} {98, 043526}

\bibitem[\protect\citeauthoryear{{Adhikari}, {Dalal}  \&
  {Chamberlain}}{{Adhikari} et~al.}{2014}]{2014JCAP...11..019A}
{Adhikari} S.,  {Dalal} N.,   {Chamberlain} R.~T.,  2014, \mn@doi [\jcap]
  {10.1088/1475-7516/2014/11/019}, \href
  {https://ui.adsabs.harvard.edu/abs/2014JCAP...11..019A} {2014, 019}

\bibitem[\protect\citeauthoryear{{Adhikari} et~al.,}{{Adhikari}
  et~al.}{2021}]{2021ApJ...923...37A}
{Adhikari} S.,  et~al., 2021, \mn@doi [\apj] {10.3847/1538-4357/ac0bbc}, \href
  {https://ui.adsabs.harvard.edu/abs/2021ApJ...923...37A} {923, 37}

\bibitem[\protect\citeauthoryear{{Alabi} et~al.,}{{Alabi}
  et~al.}{2017}]{2017MNRAS.468.3949A}
{Alabi} A.~B.,  et~al., 2017, \mn@doi [\mnras] {10.1093/mnras/stx678}, \href
  {https://ui.adsabs.harvard.edu/abs/2017MNRAS.468.3949A} {468, 3949}

\bibitem[\protect\citeauthoryear{{Allen}, {Evrard}  \& {Mantz}}{{Allen}
  et~al.}{2011}]{2011ARA&A..49..409A}
{Allen} S.~W.,  {Evrard} A.~E.,   {Mantz} A.~B.,  2011, \mn@doi [\araa]
  {10.1146/annurev-astro-081710-102514}, \href
  {https://ui.adsabs.harvard.edu/abs/2011ARA&A..49..409A} {49, 409}

\bibitem[\protect\citeauthoryear{{Auger}, {Treu}, {Bolton}, {Gavazzi},
  {Koopmans}, {Marshall}, {Moustakas}  \& {Burles}}{{Auger}
  et~al.}{2010}]{2010ApJ...724..511A}
{Auger} M.~W.,  {Treu} T.,  {Bolton} A.~S.,  {Gavazzi} R.,  {Koopmans}
  L.~V.~E.,  {Marshall} P.~J.,  {Moustakas} L.~A.,   {Burles} S.,  2010,
  \mn@doi [\apj] {10.1088/0004-637X/724/1/511}, \href
  {http://adsabs.harvard.edu/abs/2010ApJ...724..511A} {724, 511}

\bibitem[\protect\citeauthoryear{{Babyk}, {McNamara}, {Nulsen}, {Hogan},
  {Vantyghem}, {Russell}, {Pulido}  \& {Edge}}{{Babyk}
  et~al.}{2018}]{2018ApJ...857...32B}
{Babyk} I.~V.,  {McNamara} B.~R.,  {Nulsen} P.~E.~J.,  {Hogan} M.~T.,
  {Vantyghem} A.~N.,  {Russell} H.~R.,  {Pulido} F.~A.,   {Edge} A.~C.,  2018,
  \mn@doi [\apj] {10.3847/1538-4357/aab3c9}, \href
  {https://ui.adsabs.harvard.edu/abs/2018ApJ...857...32B} {857, 32}

\bibitem[\protect\citeauthoryear{{Barnab{\`e}}, {Czoske}, {Koopmans}, {Treu}
  \& {Bolton}}{{Barnab{\`e}} et~al.}{2011}]{2011MNRAS.415.2215B}
{Barnab{\`e}} M.,  {Czoske} O.,  {Koopmans} L.~V.~E.,  {Treu} T.,   {Bolton}
  A.~S.,  2011, \mn@doi [\mnras] {10.1111/j.1365-2966.2011.18842.x}, \href
  {http://adsabs.harvard.edu/abs/2011MNRAS.415.2215B} {415, 2215}

\bibitem[\protect\citeauthoryear{{Barnes} et~al.,}{{Barnes}
  et~al.}{2018}]{2018MNRAS.481.1809B}
{Barnes} D.~J.,  et~al., 2018, \mn@doi [\mnras] {10.1093/mnras/sty2078}, \href
  {https://ui.adsabs.harvard.edu/abs/2018MNRAS.481.1809B} {481, 1809}

\bibitem[\protect\citeauthoryear{{Barnes}, {Vogelsberger}, {Pearce}, {Pop},
  {Kannan}, {Cao}, {Kay}  \& {Hernquist}}{{Barnes}
  et~al.}{2021}]{2021MNRAS.tmp.1271B}
{Barnes} D.~J.,  {Vogelsberger} M.,  {Pearce} F.~A.,  {Pop} A.-R.,  {Kannan}
  R.,  {Cao} K.,  {Kay} S.~T.,   {Hernquist} L.,  2021, \mn@doi [\mnras]
  {10.1093/mnras/stab1276}, \href
  {https://ui.adsabs.harvard.edu/abs/2021MNRAS.tmp.1271B} {}

\bibitem[\protect\citeauthoryear{{Bartelmann} \& {Steinmetz}}{{Bartelmann} \&
  {Steinmetz}}{1996}]{1996MNRAS.283..431B}
{Bartelmann} M.,  {Steinmetz} M.,  1996, \mn@doi [\mnras]
  {10.1093/mnras/283.2.431}, \href
  {https://ui.adsabs.harvard.edu/abs/1996MNRAS.283..431B} {283, 431}

\bibitem[\protect\citeauthoryear{{Bellstedt} et~al.,}{{Bellstedt}
  et~al.}{2018}]{2018MNRAS.476.4543B}
{Bellstedt} S.,  et~al., 2018, \mn@doi [\mnras] {10.1093/mnras/sty456}, \href
  {http://adsabs.harvard.edu/abs/2018MNRAS.476.4543B} {476, 4543}

\bibitem[\protect\citeauthoryear{{Birnboim} \& {Dekel}}{{Birnboim} \&
  {Dekel}}{2003}]{2003MNRAS.345..349B}
{Birnboim} Y.,  {Dekel} A.,  2003, \mn@doi [\mnras]
  {10.1046/j.1365-8711.2003.06955.x}, \href
  {https://ui.adsabs.harvard.edu/abs/2003MNRAS.345..349B} {345, 349}

\bibitem[\protect\citeauthoryear{{Bleem} et~al.,}{{Bleem}
  et~al.}{2015}]{2015ApJS..216...27B}
{Bleem} L.~E.,  et~al., 2015, \mn@doi [\apjs] {10.1088/0067-0049/216/2/27},
  \href {https://ui.adsabs.harvard.edu/abs/2015ApJS..216...27B} {216, 27}

\bibitem[\protect\citeauthoryear{{Boroson}, {Kim}  \& {Fabbiano}}{{Boroson}
  et~al.}{2011}]{2011ApJ...729...12B}
{Boroson} B.,  {Kim} D.-W.,   {Fabbiano} G.,  2011, \mn@doi [\apj]
  {10.1088/0004-637X/729/1/12}, \href
  {https://ui.adsabs.harvard.edu/abs/2011ApJ...729...12B} {729, 12}

\bibitem[\protect\citeauthoryear{{Borrow}, {Vogelsberger}, {O'Neil}, {McDonald}
   \& {Smith}}{{Borrow} et~al.}{2023}]{2023MNRAS.520..649B}
{Borrow} J.,  {Vogelsberger} M.,  {O'Neil} S.,  {McDonald} M.~A.,   {Smith} A.,
   2023, \mn@doi [\mnras] {10.1093/mnras/stad045}, \href
  {https://ui.adsabs.harvard.edu/abs/2023MNRAS.520..649B} {520, 649}

\bibitem[\protect\citeauthoryear{{Bruzual} \& {Charlot}}{{Bruzual} \&
  {Charlot}}{2003}]{2003MNRAS.344.1000B}
{Bruzual} G.,  {Charlot} S.,  2003, \mn@doi [\mnras]
  {10.1046/j.1365-8711.2003.06897.x}, \href
  {http://adsabs.harvard.edu/abs/2003MNRAS.344.1000B} {344, 1000}

\bibitem[\protect\citeauthoryear{{Bryan} \& {Norman}}{{Bryan} \&
  {Norman}}{1998}]{1998ApJ...495...80B}
{Bryan} G.~L.,  {Norman} M.~L.,  1998, \mn@doi [\apj] {10.1086/305262}, \href
  {https://ui.adsabs.harvard.edu/abs/1998ApJ...495...80B} {495, 80}

\bibitem[\protect\citeauthoryear{{Cappellari} et~al.,}{{Cappellari}
  et~al.}{2011}]{2011MNRAS.416.1680C}
{Cappellari} M.,  et~al., 2011, \mn@doi [\mnras]
  {10.1111/j.1365-2966.2011.18600.x}, \href
  {https://ui.adsabs.harvard.edu/abs/2011MNRAS.416.1680C} {416, 1680}

\bibitem[\protect\citeauthoryear{{Casey}, {Greco}, {Peter}  \& {Davis}}{{Casey}
  et~al.}{2022}]{2022arXiv221100629C}
{Casey} K.~J.,  {Greco} J.~P.,  {Peter} A. H.~G.,   {Davis} A.~B.,  2022, arXiv
  e-prints, \href {https://ui.adsabs.harvard.edu/abs/2022arXiv221100629C} {p.
  arXiv:2211.00629}

\bibitem[\protect\citeauthoryear{{Ceverino} \& {Klypin}}{{Ceverino} \&
  {Klypin}}{2009}]{2009ApJ...695..292C}
{Ceverino} D.,  {Klypin} A.,  2009, \mn@doi [\apj]
  {10.1088/0004-637X/695/1/292}, \href
  {https://ui.adsabs.harvard.edu/abs/2009ApJ...695..292C} {695, 292}

\bibitem[\protect\citeauthoryear{{Choi}, {Ostriker}, {Naab}, {Oser}  \&
  {Moster}}{{Choi} et~al.}{2015}]{2015MNRAS.449.4105C}
{Choi} E.,  {Ostriker} J.~P.,  {Naab} T.,  {Oser} L.,   {Moster} B.~P.,  2015,
  \mn@doi [\mnras] {10.1093/mnras/stv575}, \href
  {https://ui.adsabs.harvard.edu/abs/2015MNRAS.449.4105C} {449, 4105}

\bibitem[\protect\citeauthoryear{{Ciotti}, {Pellegrini}, {Negri}  \&
  {Ostriker}}{{Ciotti} et~al.}{2017}]{2017ApJ...835...15C}
{Ciotti} L.,  {Pellegrini} S.,  {Negri} A.,   {Ostriker} J.~P.,  2017, \mn@doi
  [\apj] {10.3847/1538-4357/835/1/15}, \href
  {https://ui.adsabs.harvard.edu/abs/2017ApJ...835...15C} {835, 15}

\bibitem[\protect\citeauthoryear{{Croton} et~al.,}{{Croton}
  et~al.}{2006}]{2006MNRAS.365...11C}
{Croton} D.~J.,  et~al., 2006, \mn@doi [\mnras]
  {10.1111/j.1365-2966.2005.09675.x}, \href
  {https://ui.adsabs.harvard.edu/abs/2006MNRAS.365...11C} {365, 11}

\bibitem[\protect\citeauthoryear{{David}, {Jones}, {Forman}, {Vargas}  \&
  {Nulsen}}{{David} et~al.}{2006}]{2006ApJ...653..207D}
{David} L.~P.,  {Jones} C.,  {Forman} W.,  {Vargas} I.~M.,   {Nulsen} P.,
  2006, \mn@doi [\apj] {10.1086/508623}, \href
  {https://ui.adsabs.harvard.edu/abs/2006ApJ...653..207D} {653, 207}

\bibitem[\protect\citeauthoryear{{Derkenne}, {McDermid}, {Poci}, {Remus},
  {J{\o}rgensen}  \& {Emsellem}}{{Derkenne} et~al.}{2021}]{2021MNRAS.506.3691D}
{Derkenne} C.,  {McDermid} R.~M.,  {Poci} A.,  {Remus} R.-S.,  {J{\o}rgensen}
  I.,   {Emsellem} E.,  2021, \mn@doi [\mnras] {10.1093/mnras/stab1996}, \href
  {https://ui.adsabs.harvard.edu/abs/2021MNRAS.506.3691D} {506, 3691}

\bibitem[\protect\citeauthoryear{{Di Matteo}, {Springel}  \& {Hernquist}}{{Di
  Matteo} et~al.}{2005}]{2005Natur.433..604D}
{Di Matteo} T.,  {Springel} V.,   {Hernquist} L.,  2005, \mn@doi [\nat]
  {10.1038/nature03335}, \href
  {http://adsabs.harvard.edu/abs/2005Natur.433..604D} {433, 604}

\bibitem[\protect\citeauthoryear{{Diemer} \& {Kravtsov}}{{Diemer} \&
  {Kravtsov}}{2014}]{2014ApJ...789....1D}
{Diemer} B.,  {Kravtsov} A.~V.,  2014, \mn@doi [\apj]
  {10.1088/0004-637X/789/1/1}, \href
  {https://ui.adsabs.harvard.edu/abs/2014ApJ...789....1D} {789, 1}

\bibitem[\protect\citeauthoryear{{Dolag}, {Borgani}, {Murante}  \&
  {Springel}}{{Dolag} et~al.}{2009}]{2009MNRAS.399..497D}
{Dolag} K.,  {Borgani} S.,  {Murante} G.,   {Springel} V.,  2009, \mn@doi
  [\mnras] {10.1111/j.1365-2966.2009.15034.x}, \href
  {http://adsabs.harvard.edu/abs/2009MNRAS.399..497D} {399, 497}

\bibitem[\protect\citeauthoryear{{Donnari} et~al.,}{{Donnari}
  et~al.}{2019}]{2019MNRAS.485.4817D}
{Donnari} M.,  et~al., 2019, \mn@doi [\mnras] {10.1093/mnras/stz712}, \href
  {https://ui.adsabs.harvard.edu/abs/2019MNRAS.485.4817D} {485, 4817}

\bibitem[\protect\citeauthoryear{{Donnari} et~al.,}{{Donnari}
  et~al.}{2021}]{2021MNRAS.500.4004D}
{Donnari} M.,  et~al., 2021, \mn@doi [\mnras] {10.1093/mnras/staa3006}, \href
  {https://ui.adsabs.harvard.edu/abs/2021MNRAS.500.4004D} {500, 4004}

\bibitem[\protect\citeauthoryear{{Eisenstein} et~al.,}{{Eisenstein}
  et~al.}{2011}]{2011AJ....142...72E}
{Eisenstein} D.~J.,  et~al., 2011, \mn@doi [\aj] {10.1088/0004-6256/142/3/72},
  \href {https://ui.adsabs.harvard.edu/abs/2011AJ....142...72E} {142, 72}

\bibitem[\protect\citeauthoryear{{Etherington} et~al.,}{{Etherington}
  et~al.}{2022}]{2022arXiv220704070E}
{Etherington} A.,  et~al., 2022, arXiv e-prints, \href
  {https://ui.adsabs.harvard.edu/abs/2022arXiv220704070E} {p. arXiv:2207.04070}

\bibitem[\protect\citeauthoryear{{Fabian}}{{Fabian}}{2012}]{2012ARA&A..50..455F}
{Fabian} A.~C.,  2012, \mn@doi [\araa] {10.1146/annurev-astro-081811-125521},
  \href {http://adsabs.harvard.edu/abs/2012ARA%26A..50..455F} {50, 455}

\bibitem[\protect\citeauthoryear{{Forbes}, {Alabi}, {Romanowsky}, {Kim},
  {Brodie}  \& {Fabbiano}}{{Forbes} et~al.}{2017}]{2017MNRAS.464L..26F}
{Forbes} D.~A.,  {Alabi} A.,  {Romanowsky} A.~J.,  {Kim} D.-W.,  {Brodie}
  J.~P.,   {Fabbiano} G.,  2017, \mn@doi [\mnras] {10.1093/mnrasl/slw176},
  \href {https://ui.adsabs.harvard.edu/abs/2017MNRAS.464L..26F} {464, L26}

\bibitem[\protect\citeauthoryear{{Foster}, {Ji}, {Smith}  \&
  {Brickhouse}}{{Foster} et~al.}{2012}]{2012ApJ...756..128F}
{Foster} A.~R.,  {Ji} L.,  {Smith} R.~K.,   {Brickhouse} N.~S.,  2012, \mn@doi
  [\apj] {10.1088/0004-637X/756/2/128}, \href
  {https://ui.adsabs.harvard.edu/abs/2012ApJ...756..128F} {756, 128}

\bibitem[\protect\citeauthoryear{{Gaskin} et~al.,}{{Gaskin}
  et~al.}{2019}]{2019JATIS...5b1001G}
{Gaskin} J.~A.,  et~al., 2019, \mn@doi [Journal of Astronomical Telescopes,
  Instruments, and Systems] {10.1117/1.JATIS.5.2.021001}, \href
  {https://ui.adsabs.harvard.edu/abs/2019JATIS...5b1001G} {5, 021001}

\bibitem[\protect\citeauthoryear{{Genel} et~al.,}{{Genel}
  et~al.}{2014}]{2014MNRAS.445..175G}
{Genel} S.,  et~al., 2014, \mn@doi [\mnras] {10.1093/mnras/stu1654}, \href
  {http://adsabs.harvard.edu/abs/2014MNRAS.445..175G} {445, 175}

\bibitem[\protect\citeauthoryear{{Genel} et~al.,}{{Genel}
  et~al.}{2018}]{2018MNRAS.474.3976G}
{Genel} S.,  et~al., 2018, \mn@doi [\mnras] {10.1093/mnras/stx3078}, \href
  {http://adsabs.harvard.edu/abs/2018MNRAS.474.3976G} {474, 3976}

\bibitem[\protect\citeauthoryear{{Goulding} et~al.,}{{Goulding}
  et~al.}{2016}]{2016ApJ...826..167G}
{Goulding} A.~D.,  et~al., 2016, \mn@doi [\apj] {10.3847/0004-637X/826/2/167},
  \href {https://ui.adsabs.harvard.edu/abs/2016ApJ...826..167G} {826, 167}

\bibitem[\protect\citeauthoryear{{Habouzit} et~al.,}{{Habouzit}
  et~al.}{2019}]{2019MNRAS.484.4413H}
{Habouzit} M.,  et~al., 2019, \mn@doi [\mnras] {10.1093/mnras/stz102}, \href
  {https://ui.adsabs.harvard.edu/abs/2019MNRAS.484.4413H} {484, 4413}

\bibitem[\protect\citeauthoryear{{Harris}, {Harris}  \& {Alessi}}{{Harris}
  et~al.}{2013}]{2013ApJ...772...82H}
{Harris} W.~E.,  {Harris} G. L.~H.,   {Alessi} M.,  2013, \mn@doi [\apj]
  {10.1088/0004-637X/772/2/82}, \href
  {https://ui.adsabs.harvard.edu/abs/2013ApJ...772...82H} {772, 82}

\bibitem[\protect\citeauthoryear{{Harris}, {Blakeslee}  \& {Harris}}{{Harris}
  et~al.}{2017}]{2017ApJ...836...67H}
{Harris} W.~E.,  {Blakeslee} J.~P.,   {Harris} G. L.~H.,  2017, \mn@doi [\apj]
  {10.3847/1538-4357/836/1/67}, \href
  {https://ui.adsabs.harvard.edu/abs/2017ApJ...836...67H} {836, 67}

\bibitem[\protect\citeauthoryear{{Hemler} et~al.,}{{Hemler}
  et~al.}{2021}]{2021MNRAS.506.3024H}
{Hemler} Z.~S.,  et~al., 2021, \mn@doi [\mnras] {10.1093/mnras/stab1803}, \href
  {https://ui.adsabs.harvard.edu/abs/2021MNRAS.506.3024H} {506, 3024}

\bibitem[\protect\citeauthoryear{{Hilton} et~al.,}{{Hilton}
  et~al.}{2018}]{2018ApJS..235...20H}
{Hilton} M.,  et~al., 2018, \mn@doi [\apjs] {10.3847/1538-4365/aaa6cb}, \href
  {https://ui.adsabs.harvard.edu/abs/2018ApJS..235...20H} {235, 20}

\bibitem[\protect\citeauthoryear{{Hoekstra}, {Herbonnet}, {Muzzin}, {Babul},
  {Mahdavi}, {Viola}  \& {Cacciato}}{{Hoekstra}
  et~al.}{2015}]{2015MNRAS.449..685H}
{Hoekstra} H.,  {Herbonnet} R.,  {Muzzin} A.,  {Babul} A.,  {Mahdavi} A.,
  {Viola} M.,   {Cacciato} M.,  2015, \mn@doi [\mnras] {10.1093/mnras/stv275},
  \href {https://ui.adsabs.harvard.edu/abs/2015MNRAS.449..685H} {449, 685}

\bibitem[\protect\citeauthoryear{{Kauffmann}, {Nelson}, {Borthakur}, {Heckman},
  {Hernquist}, {Marinacci}, {Pakmor}  \& {Pillepich}}{{Kauffmann}
  et~al.}{2019}]{2019MNRAS.486.4686K}
{Kauffmann} G.,  {Nelson} D.,  {Borthakur} S.,  {Heckman} T.,  {Hernquist} L.,
  {Marinacci} F.,  {Pakmor} R.,   {Pillepich} A.,  2019, \mn@doi [\mnras]
  {10.1093/mnras/stz1029}, \href
  {https://ui.adsabs.harvard.edu/abs/2019MNRAS.486.4686K} {486, 4686}

\bibitem[\protect\citeauthoryear{{Kim} \& {Fabbiano}}{{Kim} \&
  {Fabbiano}}{2013}]{2013ApJ...776..116K}
{Kim} D.-W.,  {Fabbiano} G.,  2013, \mn@doi [\apj]
  {10.1088/0004-637X/776/2/116}, \href
  {https://ui.adsabs.harvard.edu/abs/2013ApJ...776..116K} {776, 116}

\bibitem[\protect\citeauthoryear{{Kim} \& {Fabbiano}}{{Kim} \&
  {Fabbiano}}{2015}]{2015ApJ...812..127K}
{Kim} D.-W.,  {Fabbiano} G.,  2015, \mn@doi [\apj]
  {10.1088/0004-637X/812/2/127}, \href
  {https://ui.adsabs.harvard.edu/abs/2015ApJ...812..127K} {812, 127}

\bibitem[\protect\citeauthoryear{{Kim} et~al.,}{{Kim}
  et~al.}{2019a}]{2019ApJS..241...36K}
{Kim} D.-W.,  et~al., 2019a, \mn@doi [\apjs] {10.3847/1538-4365/ab0ca4}, \href
  {https://ui.adsabs.harvard.edu/abs/2019ApJS..241...36K} {241, 36}

\bibitem[\protect\citeauthoryear{{Kim}, {James}, {Fabbiano}, {Forbes}  \&
  {Alabi}}{{Kim} et~al.}{2019b}]{2019MNRAS.488.1072K}
{Kim} D.-W.,  {James} N.,  {Fabbiano} G.,  {Forbes} D.,   {Alabi} A.,  2019b,
  \mn@doi [\mnras] {10.1093/mnras/stz1522}, \href
  {https://ui.adsabs.harvard.edu/abs/2019MNRAS.488.1072K} {488, 1072}

\bibitem[\protect\citeauthoryear{{Kim} et~al.,}{{Kim}
  et~al.}{2020}]{2020MNRAS.492.2095K}
{Kim} D.-W.,  et~al., 2020, \mn@doi [\mnras] {10.1093/mnras/stz3530}, \href
  {https://ui.adsabs.harvard.edu/abs/2020MNRAS.492.2095K} {492, 2095}

\bibitem[\protect\citeauthoryear{{Kormendy}, {Fisher}, {Cornell}  \&
  {Bender}}{{Kormendy} et~al.}{2009}]{2009ApJS..182..216K}
{Kormendy} J.,  {Fisher} D.~B.,  {Cornell} M.~E.,   {Bender} R.,  2009, \mn@doi
  [\apjs] {10.1088/0067-0049/182/1/216}, \href
  {http://adsabs.harvard.edu/abs/2009ApJS..182..216K} {182, 216}

\bibitem[\protect\citeauthoryear{{Kravtsov} \& {Borgani}}{{Kravtsov} \&
  {Borgani}}{2012}]{2012ARA&A..50..353K}
{Kravtsov} A.~V.,  {Borgani} S.,  2012, \mn@doi [\araa]
  {10.1146/annurev-astro-081811-125502}, \href
  {https://ui.adsabs.harvard.edu/abs/2012ARA&A..50..353K} {50, 353}

\bibitem[\protect\citeauthoryear{{Lee} et~al.,}{{Lee}
  et~al.}{2022}]{2022MNRAS.517.5303L}
{Lee} E.,  et~al., 2022, \mn@doi [\mnras] {10.1093/mnras/stac2781}, \href
  {https://ui.adsabs.harvard.edu/abs/2022MNRAS.517.5303L} {517, 5303}

\bibitem[\protect\citeauthoryear{{Li}, {Wang}, {Mo}, {Wang}, {Luo}  \&
  {Li}}{{Li} et~al.}{2022}]{2022arXiv220907711L}
{Li} H.,  {Wang} H.,  {Mo} H.~J.,  {Wang} Y.,  {Luo} X.,   {Li} R.,  2022,
  arXiv e-prints, \href {https://ui.adsabs.harvard.edu/abs/2022arXiv220907711L}
  {p. arXiv:2209.07711}

\bibitem[\protect\citeauthoryear{{Lovell} et~al.,}{{Lovell}
  et~al.}{2018}]{2018MNRAS.481.1950L}
{Lovell} M.~R.,  et~al., 2018, \mn@doi [\mnras] {10.1093/mnras/sty2339}, \href
  {http://adsabs.harvard.edu/abs/2018MNRAS.481.1950L} {481, 1950}

\bibitem[\protect\citeauthoryear{{Ma}, {Greene}, {McConnell}, {Janish},
  {Blakeslee}, {Thomas}  \& {Murphy}}{{Ma} et~al.}{2014}]{2014ApJ...795..158M}
{Ma} C.-P.,  {Greene} J.~E.,  {McConnell} N.,  {Janish} R.,  {Blakeslee} J.~P.,
   {Thomas} J.,   {Murphy} J.~D.,  2014, \mn@doi [\apj]
  {10.1088/0004-637X/795/2/158}, \href
  {http://adsabs.harvard.edu/abs/2014ApJ...795..158M} {795, 158}

\bibitem[\protect\citeauthoryear{{Mansfield}, {Kravtsov}  \&
  {Diemer}}{{Mansfield} et~al.}{2017}]{2017ApJ...841...34M}
{Mansfield} P.,  {Kravtsov} A.~V.,   {Diemer} B.,  2017, \mn@doi [\apj]
  {10.3847/1538-4357/aa7047}, \href
  {https://ui.adsabs.harvard.edu/abs/2017ApJ...841...34M} {841, 34}

\bibitem[\protect\citeauthoryear{{Mantz}, {Allen}, {Rapetti}  \&
  {Ebeling}}{{Mantz} et~al.}{2010a}]{2010MNRAS.406.1759M}
{Mantz} A.,  {Allen} S.~W.,  {Rapetti} D.,   {Ebeling} H.,  2010a, \mn@doi
  [\mnras] {10.1111/j.1365-2966.2010.16992.x}, \href
  {https://ui.adsabs.harvard.edu/abs/2010MNRAS.406.1759M} {406, 1759}

\bibitem[\protect\citeauthoryear{{Mantz}, {Allen}, {Ebeling}, {Rapetti}  \&
  {Drlica-Wagner}}{{Mantz} et~al.}{2010b}]{2010MNRAS.406.1773M}
{Mantz} A.,  {Allen} S.~W.,  {Ebeling} H.,  {Rapetti} D.,   {Drlica-Wagner} A.,
   2010b, \mn@doi [\mnras] {10.1111/j.1365-2966.2010.16993.x}, \href
  {https://ui.adsabs.harvard.edu/abs/2010MNRAS.406.1773M} {406, 1773}

\bibitem[\protect\citeauthoryear{{Mantz} et~al.,}{{Mantz}
  et~al.}{2015}]{2015MNRAS.446.2205M}
{Mantz} A.~B.,  et~al., 2015, \mn@doi [\mnras] {10.1093/mnras/stu2096}, \href
  {https://ui.adsabs.harvard.edu/abs/2015MNRAS.446.2205M} {446, 2205}

\bibitem[\protect\citeauthoryear{{Mantz} et~al.,}{{Mantz}
  et~al.}{2016}]{2016MNRAS.463.3582M}
{Mantz} A.~B.,  et~al., 2016, \mn@doi [\mnras] {10.1093/mnras/stw2250}, \href
  {https://ui.adsabs.harvard.edu/abs/2016MNRAS.463.3582M} {463, 3582}

\bibitem[\protect\citeauthoryear{{Marinacci} et~al.,}{{Marinacci}
  et~al.}{2018}]{2018MNRAS.480.5113M}
{Marinacci} F.,  et~al., 2018, \mn@doi [\mnras] {10.1093/mnras/sty2206}, \href
  {http://adsabs.harvard.edu/abs/2018MNRAS.480.5113M} {480, 5113}

\bibitem[\protect\citeauthoryear{{Mathiesen} \& {Evrard}}{{Mathiesen} \&
  {Evrard}}{2001}]{2001ApJ...546..100M}
{Mathiesen} B.~F.,  {Evrard} A.~E.,  2001, \mn@doi [\apj] {10.1086/318249},
  \href {https://ui.adsabs.harvard.edu/abs/2001ApJ...546..100M} {546, 100}

\bibitem[\protect\citeauthoryear{{Mazzotta}, {Rasia}, {Moscardini}  \&
  {Tormen}}{{Mazzotta} et~al.}{2004}]{2004MNRAS.354...10M}
{Mazzotta} P.,  {Rasia} E.,  {Moscardini} L.,   {Tormen} G.,  2004, \mn@doi
  [\mnras] {10.1111/j.1365-2966.2004.08167.x}, \href
  {https://ui.adsabs.harvard.edu/abs/2004MNRAS.354...10M} {354, 10}

\bibitem[\protect\citeauthoryear{{Merloni} et~al.,}{{Merloni}
  et~al.}{2012}]{2012arXiv1209.3114M}
{Merloni} A.,  et~al., 2012, arXiv e-prints, \href
  {https://ui.adsabs.harvard.edu/abs/2012arXiv1209.3114M} {p. arXiv:1209.3114}

\bibitem[\protect\citeauthoryear{{More}, {Diemer}  \& {Kravtsov}}{{More}
  et~al.}{2015}]{2015ApJ...810...36M}
{More} S.,  {Diemer} B.,   {Kravtsov} A.~V.,  2015, \mn@doi [\apj]
  {10.1088/0004-637X/810/1/36}, \href
  {https://ui.adsabs.harvard.edu/abs/2015ApJ...810...36M} {810, 36}

\bibitem[\protect\citeauthoryear{{More} et~al.,}{{More}
  et~al.}{2016}]{2016ApJ...825...39M}
{More} S.,  et~al., 2016, \mn@doi [\apj] {10.3847/0004-637X/825/1/39}, \href
  {https://ui.adsabs.harvard.edu/abs/2016ApJ...825...39M} {825, 39}

\bibitem[\protect\citeauthoryear{{Naiman} et~al.,}{{Naiman}
  et~al.}{2018}]{2018MNRAS.477.1206N}
{Naiman} J.~P.,  et~al., 2018, \mn@doi [\mnras] {10.1093/mnras/sty618}, \href
  {http://adsabs.harvard.edu/abs/2018MNRAS.477.1206N} {477, 1206}

\bibitem[\protect\citeauthoryear{{Nandra} et~al.,}{{Nandra}
  et~al.}{2013}]{2013arXiv1306.2307N}
{Nandra} K.,  et~al., 2013, arXiv e-prints, \href
  {https://ui.adsabs.harvard.edu/abs/2013arXiv1306.2307N} {p. arXiv:1306.2307}

\bibitem[\protect\citeauthoryear{{Nelson} et~al.,}{{Nelson}
  et~al.}{2015}]{2015A&C....13...12N}
{Nelson} D.,  et~al., 2015, \mn@doi [Astronomy and Computing]
  {10.1016/j.ascom.2015.09.003}, \href
  {http://adsabs.harvard.edu/abs/2015A%26C....13...12N} {13, 12}

\bibitem[\protect\citeauthoryear{{Nelson} et~al.,}{{Nelson}
  et~al.}{2018}]{2018MNRAS.475..624N}
{Nelson} D.,  et~al., 2018, \mn@doi [\mnras] {10.1093/mnras/stx3040}, \href
  {http://adsabs.harvard.edu/abs/2018MNRAS.475..624N} {475, 624}

\bibitem[\protect\citeauthoryear{{Nelson} et~al.,}{{Nelson}
  et~al.}{2019a}]{2019ComAC...6....2N}
{Nelson} D.,  et~al., 2019a, \mn@doi [Computational Astrophysics and Cosmology]
  {10.1186/s40668-019-0028-x}, \href
  {https://ui.adsabs.harvard.edu/abs/2019ComAC...6....2N} {6, 2}

\bibitem[\protect\citeauthoryear{{Nelson} et~al.,}{{Nelson}
  et~al.}{2019b}]{2019MNRAS.490.3234N}
{Nelson} D.,  et~al., 2019b, \mn@doi [\mnras] {10.1093/mnras/stz2306}, \href
  {https://ui.adsabs.harvard.edu/abs/2019MNRAS.490.3234N} {490, 3234}

\bibitem[\protect\citeauthoryear{{Nelson} et~al.,}{{Nelson}
  et~al.}{2021}]{2021MNRAS.508..219N}
{Nelson} E.~J.,  et~al., 2021, \mn@doi [\mnras] {10.1093/mnras/stab2131}, \href
  {https://ui.adsabs.harvard.edu/abs/2021MNRAS.508..219N} {508, 219}

\bibitem[\protect\citeauthoryear{{Neto} et~al.,}{{Neto}
  et~al.}{2007}]{2007MNRAS.381.1450N}
{Neto} A.~F.,  et~al., 2007, \mn@doi [\mnras]
  {10.1111/j.1365-2966.2007.12381.x}, \href
  {http://adsabs.harvard.edu/abs/2007MNRAS.381.1450N} {381, 1450}

\bibitem[\protect\citeauthoryear{{O'Neil}, {Barnes}, {Vogelsberger}  \&
  {Diemer}}{{O'Neil} et~al.}{2021}]{2021MNRAS.504.4649O}
{O'Neil} S.,  {Barnes} D.~J.,  {Vogelsberger} M.,   {Diemer} B.,  2021, \mn@doi
  [\mnras] {10.1093/mnras/stab1221}, \href
  {https://ui.adsabs.harvard.edu/abs/2021MNRAS.504.4649O} {504, 4649}

\bibitem[\protect\citeauthoryear{{O'Sullivan}, {Forbes}  \&
  {Ponman}}{{O'Sullivan} et~al.}{2001}]{2001MNRAS.328..461O}
{O'Sullivan} E.,  {Forbes} D.~A.,   {Ponman} T.~J.,  2001, \mn@doi [\mnras]
  {10.1046/j.1365-8711.2001.04890.x}, \href
  {https://ui.adsabs.harvard.edu/abs/2001MNRAS.328..461O} {328, 461}

\bibitem[\protect\citeauthoryear{{O'Sullivan}, {Ponman}  \&
  {Collins}}{{O'Sullivan} et~al.}{2003}]{2003MNRAS.340.1375O}
{O'Sullivan} E.,  {Ponman} T.~J.,   {Collins} R.~S.,  2003, \mn@doi [\mnras]
  {10.1046/j.1365-8711.2003.06396.x}, \href
  {https://ui.adsabs.harvard.edu/abs/2003MNRAS.340.1375O} {340, 1375}

\bibitem[\protect\citeauthoryear{{Oppenheimer} et~al.,}{{Oppenheimer}
  et~al.}{2020}]{2020ApJ...893L..24O}
{Oppenheimer} B.~D.,  et~al., 2020, \mn@doi [\apjl] {10.3847/2041-8213/ab846f},
  \href {https://ui.adsabs.harvard.edu/abs/2020ApJ...893L..24O} {893, L24}

\bibitem[\protect\citeauthoryear{{Pellegrini}}{{Pellegrini}}{1999}]{1999A&A...351..487P}
{Pellegrini} S.,  1999, \aap, \href
  {https://ui.adsabs.harvard.edu/abs/1999A&A...351..487P} {351, 487}

\bibitem[\protect\citeauthoryear{{Pellegrini}}{{Pellegrini}}{2011}]{2011ApJ...738...57P}
{Pellegrini} S.,  2011, \mn@doi [\apj] {10.1088/0004-637X/738/1/57}, \href
  {https://ui.adsabs.harvard.edu/abs/2011ApJ...738...57P} {738, 57}

\bibitem[\protect\citeauthoryear{{Pillepich} et~al.,}{{Pillepich}
  et~al.}{2018a}]{2018MNRAS.473.4077P}
{Pillepich} A.,  et~al., 2018a, \mn@doi [\mnras] {10.1093/mnras/stx2656}, \href
  {http://adsabs.harvard.edu/abs/2018MNRAS.473.4077P} {473, 4077}

\bibitem[\protect\citeauthoryear{{Pillepich} et~al.,}{{Pillepich}
  et~al.}{2018b}]{2018MNRAS.475..648P}
{Pillepich} A.,  et~al., 2018b, \mn@doi [\mnras] {10.1093/mnras/stx3112}, \href
  {http://adsabs.harvard.edu/abs/2018MNRAS.475..648P} {475, 648}

\bibitem[\protect\citeauthoryear{{Pillepich} et~al.,}{{Pillepich}
  et~al.}{2019}]{2019MNRAS.490.3196P}
{Pillepich} A.,  et~al., 2019, \mn@doi [\mnras] {10.1093/mnras/stz2338}, \href
  {https://ui.adsabs.harvard.edu/abs/2019MNRAS.490.3196P} {490, 3196}

\bibitem[\protect\citeauthoryear{{Piotrowska}, {Bluck}, {Maiolino}  \&
  {Peng}}{{Piotrowska} et~al.}{2022}]{2022MNRAS.512.1052P}
{Piotrowska} J.~M.,  {Bluck} A. F.~L.,  {Maiolino} R.,   {Peng} Y.,  2022,
  \mn@doi [\mnras] {10.1093/mnras/stab3673}, \href
  {https://ui.adsabs.harvard.edu/abs/2022MNRAS.512.1052P} {512, 1052}

\bibitem[\protect\citeauthoryear{{Planck Collaboration} et~al.,}{{Planck
  Collaboration} et~al.}{2016a}]{2016A&A...594A..13P}
{Planck Collaboration} et~al., 2016a, \mn@doi [\aap]
  {10.1051/0004-6361/201525830}, \href
  {http://adsabs.harvard.edu/abs/2016A%26A...594A..13P} {594, A13}

\bibitem[\protect\citeauthoryear{{Planck Collaboration} et~al.,}{{Planck
  Collaboration} et~al.}{2016b}]{2016A&A...594A..27P}
{Planck Collaboration} et~al., 2016b, \mn@doi [\aap]
  {10.1051/0004-6361/201525823}, \href
  {https://ui.adsabs.harvard.edu/abs/2016A&A...594A..27P} {594, A27}

\bibitem[\protect\citeauthoryear{{Poci}, {Cappellari}  \& {McDermid}}{{Poci}
  et~al.}{2017}]{2017MNRAS.467.1397P}
{Poci} A.,  {Cappellari} M.,   {McDermid} R.~M.,  2017, \mn@doi [\mnras]
  {10.1093/mnras/stx101}, \href
  {http://adsabs.harvard.edu/abs/2017MNRAS.467.1397P} {467, 1397}

\bibitem[\protect\citeauthoryear{{Pop} et~al.,}{{Pop}
  et~al.}{2022}]{2022arXiv220511528P}
{Pop} A.-R.,  et~al., 2022, arXiv e-prints, \href
  {https://ui.adsabs.harvard.edu/abs/2022arXiv220511528P} {p. arXiv:2205.11528}

\bibitem[\protect\citeauthoryear{{Remus}, {Dolag}, {Naab}, {Burkert},
  {Hirschmann}, {Hoffmann}  \& {Johansson}}{{Remus}
  et~al.}{2017}]{2017MNRAS.464.3742R}
{Remus} R.-S.,  {Dolag} K.,  {Naab} T.,  {Burkert} A.,  {Hirschmann} M.,
  {Hoffmann} T.~L.,   {Johansson} P.~H.,  2017, \mn@doi [\mnras]
  {10.1093/mnras/stw2594}, \href
  {http://adsabs.harvard.edu/abs/2017MNRAS.464.3742R} {464, 3742}

\bibitem[\protect\citeauthoryear{{Rodriguez-Gomez} et~al.,}{{Rodriguez-Gomez}
  et~al.}{2019}]{2019MNRAS.483.4140R}
{Rodriguez-Gomez} V.,  et~al., 2019, \mn@doi [\mnras] {10.1093/mnras/sty3345},
  \href {http://adsabs.harvard.edu/abs/2019MNRAS.483.4140R} {483, 4140}

\bibitem[\protect\citeauthoryear{{Ruff}, {Gavazzi}, {Marshall}, {Treu}, {Auger}
   \& {Brault}}{{Ruff} et~al.}{2011}]{2011ApJ...727...96R}
{Ruff} A.~J.,  {Gavazzi} R.,  {Marshall} P.~J.,  {Treu} T.,  {Auger} M.~W.,
  {Brault} F.,  2011, \mn@doi [\apj] {10.1088/0004-637X/727/2/96}, \href
  {http://adsabs.harvard.edu/abs/2011ApJ...727...96R} {727, 96}

\bibitem[\protect\citeauthoryear{{Rykoff} et~al.,}{{Rykoff}
  et~al.}{2014}]{2014ApJ...785..104R}
{Rykoff} E.~S.,  et~al., 2014, \mn@doi [\apj] {10.1088/0004-637X/785/2/104},
  \href {https://ui.adsabs.harvard.edu/abs/2014ApJ...785..104R} {785, 104}

\bibitem[\protect\citeauthoryear{{Rykoff} et~al.,}{{Rykoff}
  et~al.}{2016}]{2016ApJS..224....1R}
{Rykoff} E.~S.,  et~al., 2016, \mn@doi [\apjs] {10.3847/0067-0049/224/1/1},
  \href {https://ui.adsabs.harvard.edu/abs/2016ApJS..224....1R} {224, 1}

\bibitem[\protect\citeauthoryear{{Sarzi} et~al.,}{{Sarzi}
  et~al.}{2013}]{2013MNRAS.432.1845S}
{Sarzi} M.,  et~al., 2013, \mn@doi [\mnras] {10.1093/mnras/stt062}, \href
  {https://ui.adsabs.harvard.edu/abs/2013MNRAS.432.1845S} {432, 1845}

\bibitem[\protect\citeauthoryear{{Shajib}, {Treu}, {Birrer}  \&
  {Sonnenfeld}}{{Shajib} et~al.}{2021}]{2021MNRAS.503.2380S}
{Shajib} A.~J.,  {Treu} T.,  {Birrer} S.,   {Sonnenfeld} A.,  2021, \mn@doi
  [\mnras] {10.1093/mnras/stab536}, \href
  {https://ui.adsabs.harvard.edu/abs/2021MNRAS.503.2380S} {503, 2380}

\bibitem[\protect\citeauthoryear{{Sijacki}, {Vogelsberger}, {Genel},
  {Springel}, {Torrey}, {Snyder}, {Nelson}  \& {Hernquist}}{{Sijacki}
  et~al.}{2015}]{2015MNRAS.452..575S}
{Sijacki} D.,  {Vogelsberger} M.,  {Genel} S.,  {Springel} V.,  {Torrey} P.,
  {Snyder} G.~F.,  {Nelson} D.,   {Hernquist} L.,  2015, \mn@doi [\mnras]
  {10.1093/mnras/stv1340}, \href
  {http://adsabs.harvard.edu/abs/2015MNRAS.452..575S} {452, 575}

\bibitem[\protect\citeauthoryear{{Smith}, {Brickhouse}, {Liedahl}  \&
  {Raymond}}{{Smith} et~al.}{2001}]{2001ApJ...556L..91S}
{Smith} R.~K.,  {Brickhouse} N.~S.,  {Liedahl} D.~A.,   {Raymond} J.~C.,  2001,
  \mn@doi [\apjl] {10.1086/322992}, \href
  {https://ui.adsabs.harvard.edu/abs/2001ApJ...556L..91S} {556, L91}

\bibitem[\protect\citeauthoryear{{Sonnenfeld}, {Treu}, {Gavazzi}, {Suyu},
  {Marshall}, {Auger}  \& {Nipoti}}{{Sonnenfeld}
  et~al.}{2013}]{2013ApJ...777...98S}
{Sonnenfeld} A.,  {Treu} T.,  {Gavazzi} R.,  {Suyu} S.~H.,  {Marshall} P.~J.,
  {Auger} M.~W.,   {Nipoti} C.,  2013, \mn@doi [\apj]
  {10.1088/0004-637X/777/2/98}, \href
  {http://adsabs.harvard.edu/abs/2013ApJ...777...98S} {777, 98}

\bibitem[\protect\citeauthoryear{{Springel}}{{Springel}}{2010}]{2010MNRAS.401..791S}
{Springel} V.,  2010, \mn@doi [\mnras] {10.1111/j.1365-2966.2009.15715.x},
  \href {http://adsabs.harvard.edu/abs/2010MNRAS.401..791S} {401, 791}

\bibitem[\protect\citeauthoryear{{Springel} \& {Hernquist}}{{Springel} \&
  {Hernquist}}{2003}]{2003MNRAS.339..289S}
{Springel} V.,  {Hernquist} L.,  2003, \mn@doi [\mnras]
  {10.1046/j.1365-8711.2003.06206.x}, \href
  {http://adsabs.harvard.edu/abs/2003MNRAS.339..289S} {339, 289}

\bibitem[\protect\citeauthoryear{{Springel}, {White}, {Tormen}  \&
  {Kauffmann}}{{Springel} et~al.}{2001}]{2001MNRAS.328..726S}
{Springel} V.,  {White} S.~D.~M.,  {Tormen} G.,   {Kauffmann} G.,  2001,
  \mn@doi [\mnras] {10.1046/j.1365-8711.2001.04912.x}, \href
  {http://adsabs.harvard.edu/abs/2001MNRAS.328..726S} {328, 726}

\bibitem[\protect\citeauthoryear{{Springel}, {Di Matteo}  \&
  {Hernquist}}{{Springel} et~al.}{2005}]{2005MNRAS.361..776S}
{Springel} V.,  {Di Matteo} T.,   {Hernquist} L.,  2005, \mn@doi [\mnras]
  {10.1111/j.1365-2966.2005.09238.x}, \href
  {http://adsabs.harvard.edu/abs/2005MNRAS.361..776S} {361, 776}

\bibitem[\protect\citeauthoryear{{Springel} et~al.,}{{Springel}
  et~al.}{2018}]{2018MNRAS.475..676S}
{Springel} V.,  et~al., 2018, \mn@doi [\mnras] {10.1093/mnras/stx3304}, \href
  {http://adsabs.harvard.edu/abs/2018MNRAS.475..676S} {475, 676}

\bibitem[\protect\citeauthoryear{{Terrazas} et~al.,}{{Terrazas}
  et~al.}{2020}]{2020MNRAS.493.1888T}
{Terrazas} B.~A.,  et~al., 2020, \mn@doi [\mnras] {10.1093/mnras/staa374},
  \href {https://ui.adsabs.harvard.edu/abs/2020MNRAS.493.1888T} {493, 1888}

\bibitem[\protect\citeauthoryear{{Thomas}, {Saglia}, {Bender}, {Thomas},
  {Gebhardt}, {Magorrian}, {Corsini}  \& {Wegner}}{{Thomas}
  et~al.}{2007}]{2007MNRAS.382..657T}
{Thomas} J.,  {Saglia} R.~P.,  {Bender} R.,  {Thomas} D.,  {Gebhardt} K.,
  {Magorrian} J.,  {Corsini} E.~M.,   {Wegner} G.,  2007, \mn@doi [\mnras]
  {10.1111/j.1365-2966.2007.12434.x}, \href
  {http://adsabs.harvard.edu/abs/2007MNRAS.382..657T} {382, 657}

\bibitem[\protect\citeauthoryear{{Torrey}, {Vogelsberger}, {Genel}, {Sijacki},
  {Springel}  \& {Hernquist}}{{Torrey} et~al.}{2014}]{2014MNRAS.438.1985T}
{Torrey} P.,  {Vogelsberger} M.,  {Genel} S.,  {Sijacki} D.,  {Springel} V.,
  {Hernquist} L.,  2014, \mn@doi [\mnras] {10.1093/mnras/stt2295}, \href
  {http://adsabs.harvard.edu/abs/2014MNRAS.438.1985T} {438, 1985}

\bibitem[\protect\citeauthoryear{{Torrey} et~al.,}{{Torrey}
  et~al.}{2018}]{2018MNRAS.477L..16T}
{Torrey} P.,  et~al., 2018, \mn@doi [\mnras] {10.1093/mnrasl/sly031}, \href
  {http://adsabs.harvard.edu/abs/2018MNRAS.477L..16T} {477, L16}

\bibitem[\protect\citeauthoryear{{Torrey} et~al.,}{{Torrey}
  et~al.}{2019}]{2019MNRAS.484.5587T}
{Torrey} P.,  et~al., 2019, \mn@doi [\mnras] {10.1093/mnras/stz243}, \href
  {https://ui.adsabs.harvard.edu/abs/2019MNRAS.484.5587T} {484, 5587}

\bibitem[\protect\citeauthoryear{{Tortora}, {La Barbera}, {Napolitano},
  {Romanowsky}, {Ferreras}  \& {de Carvalho}}{{Tortora}
  et~al.}{2014}]{2014MNRAS.445..115T}
{Tortora} C.,  {La Barbera} F.,  {Napolitano} N.~R.,  {Romanowsky} A.~J.,
  {Ferreras} I.,   {de Carvalho} R.~R.,  2014, \mn@doi [\mnras]
  {10.1093/mnras/stu1616}, \href
  {http://adsabs.harvard.edu/abs/2014MNRAS.445..115T} {445, 115}

\bibitem[\protect\citeauthoryear{{Truong} et~al.,}{{Truong}
  et~al.}{2020}]{2020MNRAS.494..549T}
{Truong} N.,  et~al., 2020, \mn@doi [\mnras] {10.1093/mnras/staa685}, \href
  {https://ui.adsabs.harvard.edu/abs/2020MNRAS.494..549T} {494, 549}

\bibitem[\protect\citeauthoryear{{Truong}, {Pillepich}  \& {Werner}}{{Truong}
  et~al.}{2021}]{2021MNRAS.501.2210T}
{Truong} N.,  {Pillepich} A.,   {Werner} N.,  2021, \mn@doi [\mnras]
  {10.1093/mnras/staa3880}, \href
  {https://ui.adsabs.harvard.edu/abs/2021MNRAS.501.2210T} {501, 2210}

\bibitem[\protect\citeauthoryear{{Vikhlinin} et~al.,}{{Vikhlinin}
  et~al.}{2009}]{2009ApJ...692.1033V}
{Vikhlinin} A.,  et~al., 2009, \mn@doi [\apj] {10.1088/0004-637X/692/2/1033},
  \href {https://ui.adsabs.harvard.edu/abs/2009ApJ...692.1033V} {692, 1033}

\bibitem[\protect\citeauthoryear{{Vogelsberger}, {Genel}, {Sijacki}, {Torrey},
  {Springel}  \& {Hernquist}}{{Vogelsberger}
  et~al.}{2013}]{2013MNRAS.436.3031V}
{Vogelsberger} M.,  {Genel} S.,  {Sijacki} D.,  {Torrey} P.,  {Springel} V.,
  {Hernquist} L.,  2013, \mn@doi [\mnras] {10.1093/mnras/stt1789}, \href
  {http://adsabs.harvard.edu/abs/2013MNRAS.436.3031V} {436, 3031}

\bibitem[\protect\citeauthoryear{{Vogelsberger} et~al.,}{{Vogelsberger}
  et~al.}{2014a}]{2014MNRAS.444.1518V}
{Vogelsberger} M.,  et~al., 2014a, \mn@doi [\mnras] {10.1093/mnras/stu1536},
  \href {http://adsabs.harvard.edu/abs/2014MNRAS.444.1518V} {444, 1518}

\bibitem[\protect\citeauthoryear{{Vogelsberger} et~al.,}{{Vogelsberger}
  et~al.}{2014b}]{2014Natur.509..177V}
{Vogelsberger} M.,  et~al., 2014b, \mn@doi [\nat] {10.1038/nature13316}, \href
  {http://adsabs.harvard.edu/abs/2014Natur.509..177V} {509, 177}

\bibitem[\protect\citeauthoryear{{Vogelsberger} et~al.,}{{Vogelsberger}
  et~al.}{2018}]{2018MNRAS.474.2073V}
{Vogelsberger} M.,  et~al., 2018, \mn@doi [\mnras] {10.1093/mnras/stx2955},
  \href {http://adsabs.harvard.edu/abs/2018MNRAS.474.2073V} {474, 2073}

\bibitem[\protect\citeauthoryear{{Vogelsberger}, {Marinacci}, {Torrey}  \&
  {Puchwein}}{{Vogelsberger} et~al.}{2020a}]{2020NatRP...2...42V}
{Vogelsberger} M.,  {Marinacci} F.,  {Torrey} P.,   {Puchwein} E.,  2020a,
  \mn@doi [Nature Reviews Physics] {10.1038/s42254-019-0127-2}, \href
  {https://ui.adsabs.harvard.edu/abs/2020NatRP...2...42V} {2, 42}

\bibitem[\protect\citeauthoryear{{Vogelsberger} et~al.,}{{Vogelsberger}
  et~al.}{2020b}]{2020MNRAS.492.5167V}
{Vogelsberger} M.,  et~al., 2020b, \mn@doi [\mnras] {10.1093/mnras/staa137},
  \href {https://ui.adsabs.harvard.edu/abs/2020MNRAS.492.5167V} {492, 5167}

\bibitem[\protect\citeauthoryear{{Wang} et~al.,}{{Wang}
  et~al.}{2019}]{2019MNRAS.490.5722W}
{Wang} Y.,  et~al., 2019, \mn@doi [\mnras] {10.1093/mnras/stz2907}, \href
  {https://ui.adsabs.harvard.edu/abs/2019MNRAS.490.5722W} {490, 5722}

\bibitem[\protect\citeauthoryear{{Wang} et~al.,}{{Wang}
  et~al.}{2020}]{2020MNRAS.491.5188W}
{Wang} Y.,  et~al., 2020, \mn@doi [\mnras] {10.1093/mnras/stz3348}, \href
  {https://ui.adsabs.harvard.edu/abs/2020MNRAS.491.5188W} {491, 5188}

\bibitem[\protect\citeauthoryear{{Wang}, {Mao}, {Vogelsberger}, {Springel},
  {Hernquist}  \& {Wechsler}}{{Wang} et~al.}{2022}]{2022MNRAS.513.6134W}
{Wang} Y.,  {Mao} S.,  {Vogelsberger} M.,  {Springel} V.,  {Hernquist} L.,
  {Wechsler} R.~H.,  2022, \mn@doi [\mnras] {10.1093/mnras/stac1375}, \href
  {https://ui.adsabs.harvard.edu/abs/2022MNRAS.513.6134W} {513, 6134}

\bibitem[\protect\citeauthoryear{{Weinberger} et~al.,}{{Weinberger}
  et~al.}{2017}]{2017MNRAS.465.3291W}
{Weinberger} R.,  et~al., 2017, \mn@doi [\mnras] {10.1093/mnras/stw2944}, \href
  {http://adsabs.harvard.edu/abs/2017MNRAS.465.3291W} {465, 3291}

\bibitem[\protect\citeauthoryear{{Weinberger} et~al.,}{{Weinberger}
  et~al.}{2018}]{2018MNRAS.479.4056W}
{Weinberger} R.,  et~al., 2018, \mn@doi [\mnras] {10.1093/mnras/sty1733}, \href
  {https://ui.adsabs.harvard.edu/abs/2018MNRAS.479.4056W} {479, 4056}

\bibitem[\protect\citeauthoryear{{Weinberger}, {Springel}  \&
  {Pakmor}}{{Weinberger} et~al.}{2020}]{2020ApJS..248...32W}
{Weinberger} R.,  {Springel} V.,   {Pakmor} R.,  2020, \mn@doi [\apjs]
  {10.3847/1538-4365/ab908c}, \href
  {https://ui.adsabs.harvard.edu/abs/2020ApJS..248...32W} {248, 32}

\bibitem[\protect\citeauthoryear{{White} \& {Sarazin}}{{White} \&
  {Sarazin}}{1991}]{1991ApJ...367..476W}
{White} Raymond~E. I.,  {Sarazin} C.~L.,  1991, \mn@doi [\apj]
  {10.1086/169644}, \href
  {https://ui.adsabs.harvard.edu/abs/1991ApJ...367..476W} {367, 476}

\bibitem[\protect\citeauthoryear{{Wojtak} \& {Mamon}}{{Wojtak} \&
  {Mamon}}{2013}]{2013MNRAS.428.2407W}
{Wojtak} R.,  {Mamon} G.~A.,  2013, \mn@doi [\mnras] {10.1093/mnras/sts203},
  \href {https://ui.adsabs.harvard.edu/abs/2013MNRAS.428.2407W} {428, 2407}

\bibitem[\protect\citeauthoryear{{Xu}, {Springel}, {Sluse}, {Schneider},
  {Sonnenfeld}, {Nelson}, {Vogelsberger}  \& {Hernquist}}{{Xu}
  et~al.}{2017}]{2017MNRAS.469.1824X}
{Xu} D.,  {Springel} V.,  {Sluse} D.,  {Schneider} P.,  {Sonnenfeld} A.,
  {Nelson} D.,  {Vogelsberger} M.,   {Hernquist} L.,  2017, \mn@doi [\mnras]
  {10.1093/mnras/stx899}, \href
  {http://adsabs.harvard.edu/abs/2017MNRAS.469.1824X} {469, 1824}

\bibitem[\protect\citeauthoryear{{Xu} et~al.,}{{Xu}
  et~al.}{2019}]{2019MNRAS.489..842X}
{Xu} D.,  et~al., 2019, \mn@doi [\mnras] {10.1093/mnras/stz2164}, \href
  {https://ui.adsabs.harvard.edu/abs/2019MNRAS.489..842X} {489, 842}

\bibitem[\protect\citeauthoryear{{Zinger} et~al.,}{{Zinger}
  et~al.}{2020}]{2020MNRAS.499..768Z}
{Zinger} E.,  et~al., 2020, \mn@doi [\mnras] {10.1093/mnras/staa2607}, \href
  {https://ui.adsabs.harvard.edu/abs/2020MNRAS.499..768Z} {499, 768}

\bibitem[\protect\citeauthoryear{{von der Linden} et~al.,}{{von der Linden}
  et~al.}{2014}]{2014MNRAS.439....2V}
{von der Linden} A.,  et~al., 2014, \mn@doi [\mnras] {10.1093/mnras/stt1945},
  \href {https://ui.adsabs.harvard.edu/abs/2014MNRAS.439....2V} {439, 2}

\makeatother
\end{thebibliography}




\appendix

\section{Dynamical state of the ETG host haloes}
\label{sec:App_A}

\begin{figure}
	\includegraphics[width=\columnwidth]{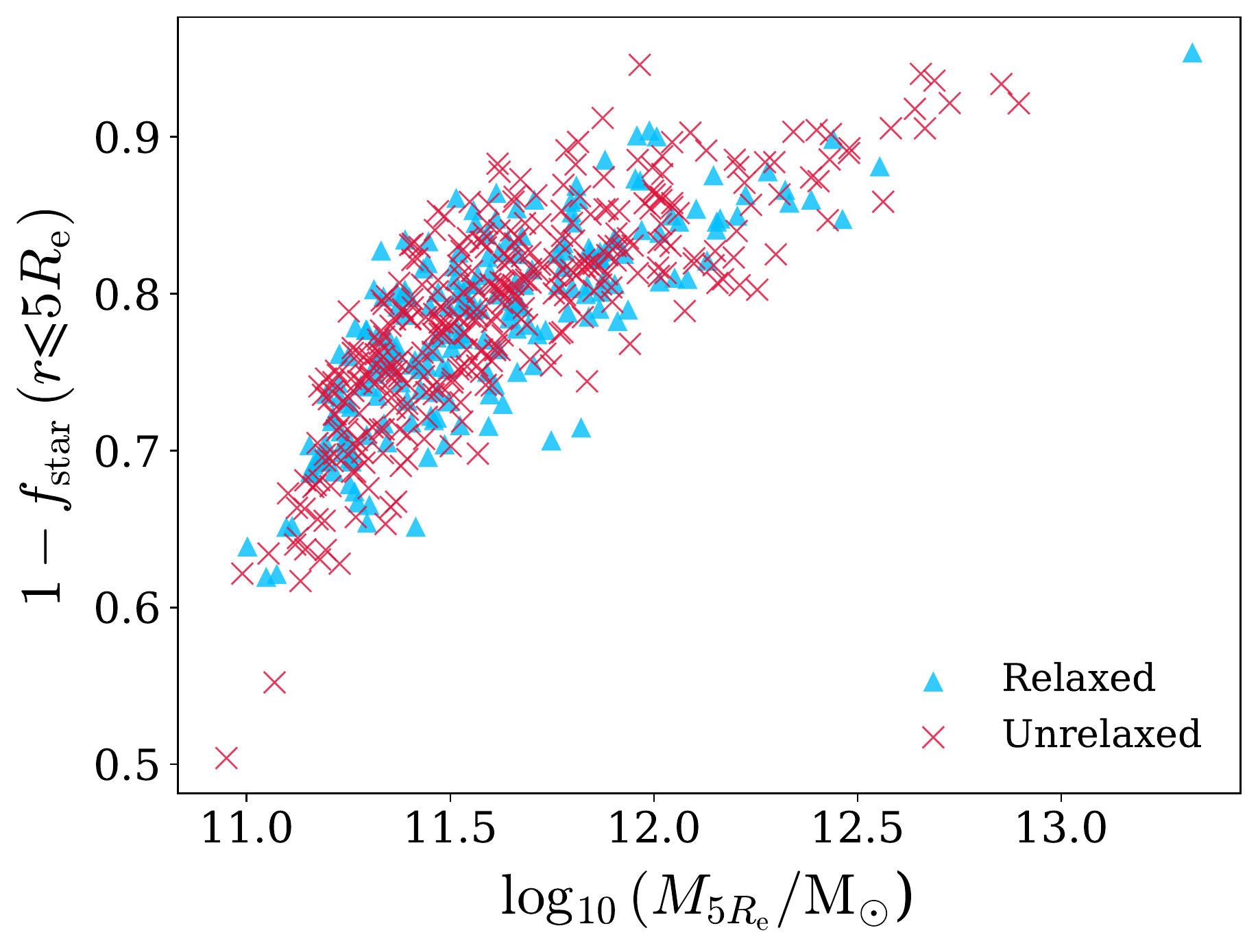}
    \caption{The dark matter fraction ($1-f_{\rm star}$) versus the total mass within $5R_{\mathrm{e}}$. The blue triangles indicate galaxies that occupy relaxed haloes, while the red crosses indicate unrelaxed host haloes. See Section~\ref{sec:App_A} for the definition of relaxation. There are no significant differences in the dark matter fraction of relaxed and unrelaxed ETG haloes.}
    \label{fig:A1}
\end{figure}

In Fig.~\ref{fig:A1}, we show the dark matter fraction ($1-f_{\rm star} (\leqslant 5R_{\rm e})$) versus the total mass of the central ETGs within $5R_{\mathrm{e}}$. The scattered dots are divided into two populations, where one denotes galaxies that occupy relaxed haloes (blue triangles), and the other denotes galaxies that live in unrelaxed ones (red crosses). The definition for relaxed haloes is adopted from~\citet{2007MNRAS.381.1450N}:
\begin{equation}
    \label{eq:A1}
    |\mathbf{r}_{\mathrm{pos}} - \mathbf{r}_{\mathrm{CM}}| \leqslant 0.07R_{200}\,,
\end{equation}
where $\mathbf{r}_{\mathrm{pos}}$ is the location of the deepest-potential particle identified in the simulation by \textsc{Subfind}, $\mathbf{r}_{\mathrm{CM}}$ is the center of mass for all particles within the FoF group, and $R_{200}$ is the physical radius within which the average density of the halo is 200 times that of the critical density of the universe. Our sample includes relaxed and unrelaxed haloes in comparable amounts according to this definition, and that they do not show significant differences in the dark matter fraction/total mass space. Hence, the dynamical state of the host halo for our ETGs should have an insignificant impact on our analysis. 

\section{Black hole mass and thermal AGN feedback effects on the luminosity-mass scaling relation}
\label{sec:App_B}

\begin{figure}
	\includegraphics[width=\columnwidth]{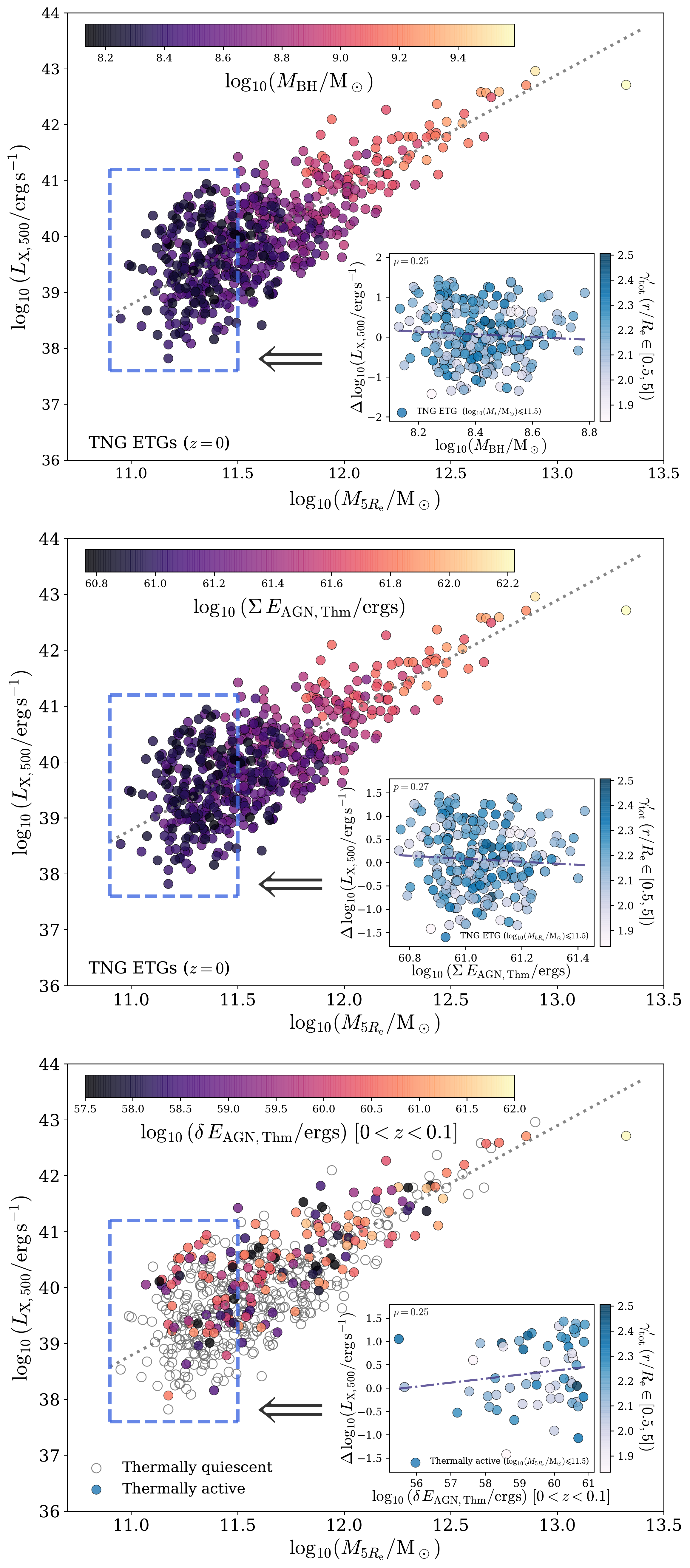}
    \caption{Similar to Fig.~\ref{fig:4}, we show the $L_{\mathrm{X, 500}}$--$M_{5R_{\rm e}}$ scaling relation colored by the central SMBH mass ($M_{\rm BH}$, {\it top panel}), cumulative radiative (thermal) mode AGN feedback energy ($\Sigma E_{\rm AGN, Thm}$, {\it middle panel}), and recent thermal AGN feedback energy within $0<z<0.1$ ($\delta E_{\rm AGN, Thm}$, {\it bottom panel}). In the bottom panel, quiescent SMBHs that did not emit any thermal energy at $z<0.1$ are denoted with empty circles.  The inset plots in each panel shows $\Delta L_{\mathrm{X, 500}}$ versus the respective colored black hole property in the main scaling relation for low-mass ETGs (blue dashed box, $\log_{10} (M_{5R_{\rm e}}/\mathrm{M_{\astrosun}}\leqslant 11.5$). The lack of color gradient in the blue boxes and large Pearson $p$ values for the linear fits in the insets  indicate insignificant correlation of $L_{\rm X, 500}$ with all three black hole properties shown in the color scale. }
    \label{fig:B}
\end{figure}

In this appendix we present the insignificant impact of SMBH mass and radiative mode AGN feedback energy on the scatter of the $L_{\mathrm{X, 500}}$--$M_{5R_{\rm e}}$ scaling relation at the low-mass end. In Fig.~\ref{fig:B}, we show the  $L_{\mathrm{X, 500}}$--$M_{5R_{\rm e}}$ scaling relation colored by the SMBH mass ($M_{\rm BH}$, top panel), cumulative thermal AGN feedback energy ($\Sigma E_{\rm AGN, Thm}$, middle panel), and the recent AGN thermal feedback energy from $z=0.1$ to $z=0$ ($\delta E_{\rm AGN, Thm}$, bottom panel). In each panel, we also include insets that further illustrate the correlation between the colored physical quantities and $\Delta L_{\rm X, 500}$ for low-mass ETGs with $\log_{10} (M_{5R_{\rm e}}/\mathrm{M_{\astrosun}}) \leqslant 11.5$. Through the color maps in the main scaling relations large Pearson $p$ values for the linear fits in the inset plots, we conclude that none of the three colored quantities show significant correlation with the scatter in $L_{\rm X}$ at the low-mass end. This indicates that the systematic variation in the {\it kinetic} mode AGN feedback energy seen in Fig.~\ref{fig:4} is not due to systematic variations in the SMBH mass, but rather intrinsic scatter in the accretion and assembly histories of those SMBHs. The middle and bottom panels of Fig.~\ref{fig:B} suggest that the scatter in $L_{\mathrm{X, 500}}$ at the low-mass end is not significantly impacted by cumulative or recent radiative mode AGN feedback, consistent with the scenario that radiative mode AGN feedback is subdominant compared to the kinetic mode in the SMBHs of $z=0$ galaxies with $M_{\ast }\gtrsim 10^{11}\,\mathrm{M_{\astrosun}}$ in IllustrisTNG~\citep{2020MNRAS.499..768Z}. Since radiative mode AGN feedback is mostly shut off for most of these ETGs below $z=0.1$ (bottom panel), it is unlikely to be an effective heat source that could alter $L_{\rm X}$ and affect the scatter in the scaling relation.

\section{Spectroscopic-like versus mass-weighted temperatures}
\label{sec:App_C}

\begin{figure}
	\includegraphics[width=\columnwidth]{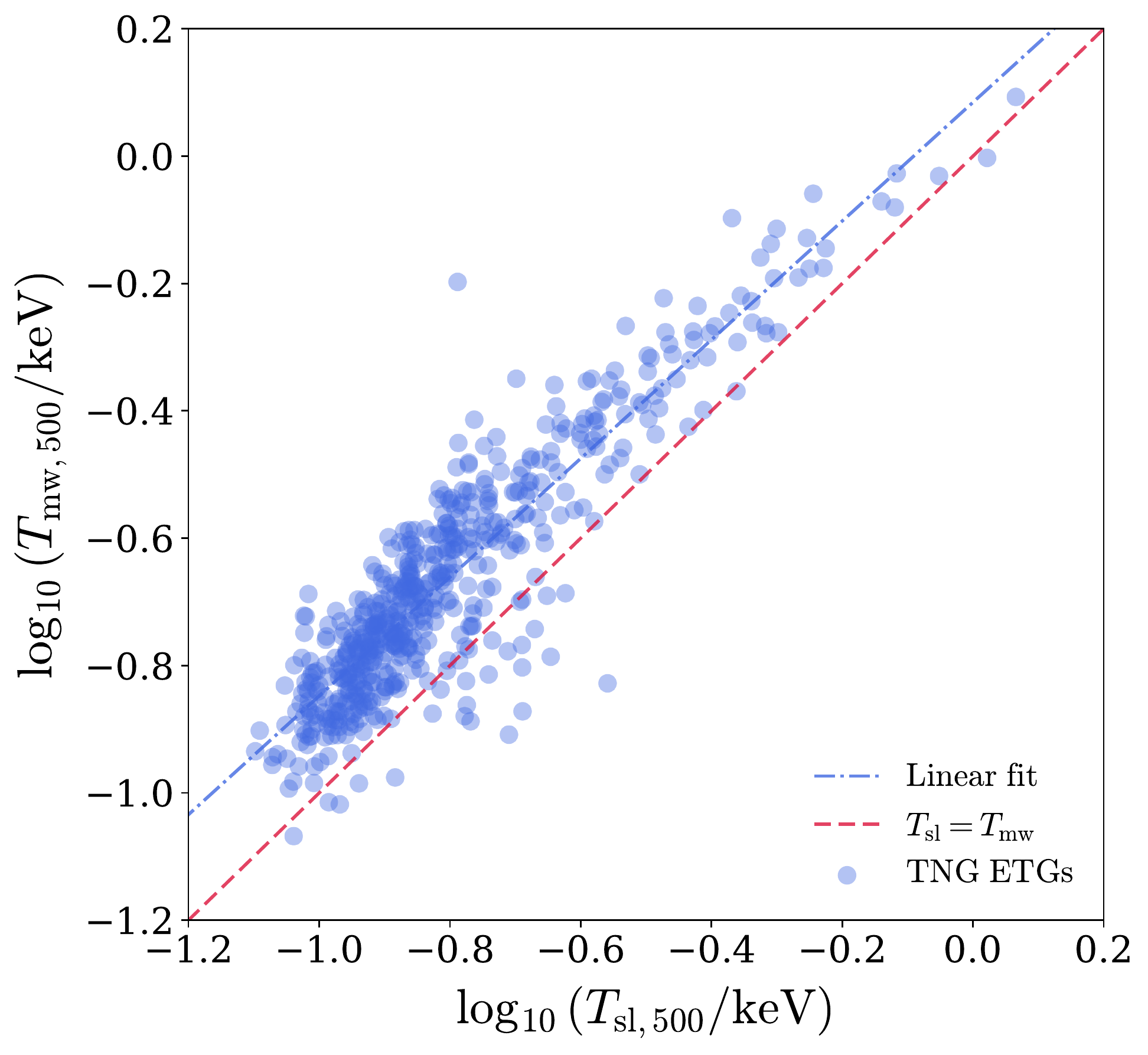}
    \caption{The spectroscopic-like temperature ($T_{\mathrm{sl, 500}}$) versus the mass-weighted temperature ($T_{\mathrm{mw, 500}}$) for our simulated ETGs (blue dots). The red dashed line stands for $T_{\mathrm{sl, 500}}=T_{\mathrm{mw, 500}}$, while the blue dotted-dashed curve is the linear fit to the log temperatures. According to the linear fit, $\log_{10}T _{\mathrm{mw, 500}}$ is mostly higher than $\log_{10} T_{\mathrm{sl, 500}}$ by $\sim 0.2$ dex.}
    \label{fig:C}
\end{figure}

In Fig.~\ref{fig:C} we show the temperature comparison between the spectroscopic-like ($T_{\mathrm{sl, 500}}$) definition and mass-weighted definition ($T_{\mathrm{mw, 500}}$). The definition of $T_{\mathrm{sl, 500}}$ follows from Equation~\ref{eq:1} and the definition of ($T_{\mathrm{mw, 500}}$) follows from Equation~\ref{eq:3}. The $T^{-3/4}$ weighting for $T_{\mathrm{sl, 500}}$ makes the $T_{\mathrm{mw, 500}}$ of the simulated ETGs $\sim 0.2$ dex higher than their $T_{\mathrm{sl, 500}}$.


\bsp	
\label{lastpage}
\end{document}